\documentclass[11pt,a4paper]{article}
\usepackage{float}
\usepackage[T1]{fontenc}
\usepackage[utf8]{inputenc}
\usepackage{lmodern}
\usepackage[a4paper,margin=25mm]{geometry}
\usepackage{amsmath,amssymb,mathtools}
\usepackage{graphicx,booktabs,pdflscape,placeins}
\usepackage{array}[=2024/01/01]
\usepackage[table]{xcolor}
\usepackage{arydshln}
\makeatletter
\def\endtabular{\endarray\m@th$\egroup}
\expandafter\let\csname endtabular*\endcsname\endtabular
\makeatother
\newcolumntype{Z}{>{\centering\arraybackslash}p{4.8em}}
\usepackage{microtype,setspace,textcomp}
\usepackage[natbibapa]{apacite}
\usepackage[font=small,labelfont=bf]{caption}
\usepackage{hyperref}

\hypersetup{pdftitle={Neural Score-Driven Volatility and Risk Forecasting},pdfsubject={GAS-NN methodology and eight-market empirical analysis}}
\definecolor{mcsgray}{gray}{0.86}
\title{Neural Score-Driven Volatility and Risk Forecasting}
\author{
Qianli Zhao\footnote{School of Business Analytics and Marketing, The University of Sydney} \and
Chao Wang\footnotemark[1]\and
Minh-Ngoc Tran\footnotemark[1]  \and
Richard Gerlach\footnotemark[1] 
}
\date{}
\begin{document}
\maketitle
\begin{abstract}

We develop a score-driven volatility model in which no particular parametric innovation distribution needs to be specified. The proposed method relies on a lightweight neural network to represent the negative score of the innovation density. By integrating and normalizing the learned score and imposing location-scale identification, we recover a flexible innovation density that can accommodate asymmetry, including negative skewness commonly observed in financial returns. Two estimation procedures are considered. A two-step procedure estimates the neural score by score matching and subsequently fits the volatility recursion with the identified density held fixed. A one-step procedure jointly refines the neural and volatility parameters under the conditional likelihood. We illustrate the practical relevance of the proposed method using eight international equity indices, with a separate evaluation of the high-volatility first half of 2020. The results show that the neural score-driven models are competitive with Gaussian and Student-$t$ GAS benchmarks that require a prespecified parametric innovation distribution, with particularly strong performance during the COVID-19 period.


\end{abstract}

\section{Introduction}
\label{introduction}

Modelling time-varying volatility is central to financial risk forecasting because the scale of return fluctuations changes over time. The ARCH and GARCH models of \citet{engle1982} and \citet{bollerslev1986} provide a parsimonious framework for capturing this variation through past squared returns and past conditional variances. For risk management, however, volatility dynamics must be considered together with the conditional return distribution, which determines the probabilities and magnitudes of extreme returns and therefore the forecasts of Value-at-Risk (VaR) and Expected Shortfall (ES).

The generalized autoregressive score (GAS) framework of \citet{creal2013}, also developed as dynamic conditional score models by \citet{harvey2013}, provides a systematic connection between the conditional distribution and time-varying parameter dynamics. In this framework, parameters are updated using the score of the conditional log likelihood, defined as its derivative with respect to the time-varying parameter, together with an appropriate scaling factor. For volatility models, the score determines how each new return updates the conditional variance. The resulting models retain a tractable observation-driven recursion and can be estimated by maximum likelihood.

In conventional GAS specifications, an innovation distribution, such as the Gaussian or Student-$t$, is specified in advance, and the corresponding score function follows from that distribution. This choice determines both the predictive return distribution and the volatility response to new observations. For example, the Gaussian log-variance score responds quadratically to standardized returns, whereas the standardized Student-$t$ log-variance score remains bounded as returns become extreme. Although these parametric specifications are parsimonious, their prescribed distributional shapes may restrict the model's ability to capture asymmetry and other features of financial returns. This motivates estimating the innovation distribution and its associated score from the data while retaining the GAS updating structure.

Semiparametric score-driven models provide a basis for this approach. \citet{blasques2016} estimate the innovation density using kernel density estimation applied to standardized residuals from an initial parametric volatility model. The estimated density determines both the conditional likelihood and the volatility update, and the procedure can alternate between updating the density and re-estimating the dynamic parameters. \citet{vanbrummelen2026} instead represent the negative innovation-density score by a natural cubic spline and reconstruct the corresponding density through integration and normalization. For a given knot vector, the spline parameters and the dynamic parameters are estimated jointly by maximum likelihood. These studies demonstrate how data-based density estimation can be incorporated into score-driven dynamics without imposing a conventional parametric innovation distribution.

The implementation of these approaches nevertheless requires choices concerning the kernel and bandwidth or the number and placement of spline knots. This motivates an alternative representation that learns the score function without specifying a kernel or a knot sequence. Feedforward neural networks provide such a representation by learning smooth nonlinear functions through adjustable weights and biases. Their flexibility can be controlled through the network architecture and regularization. In the present setting, a network can represent the negative innovation-density score, allowing its shape to be learned from the data while the GAS recursion continues to describe temporal dependence.

Score matching provides a way to estimate this neural representation. \citet{hyvarinen2005} develops an estimation method based on matching the gradient of a model's log density to that of the data density. Under suitable regularity and boundary conditions, the resulting objective can be evaluated without knowing the data density or computing the model's normalization constant. This is particularly useful when a neural network represents the density score directly. In the univariate setting considered here, integrating the estimated negative score and normalizing the result recovers an associated density, which can subsequently be used for likelihood estimation and risk forecasting.

Against this background, this paper develops GAS-NN, a neural score-driven volatility model in which a compact feedforward network represents a raw negative density score. Integration and normalization recover the associated raw density, which is then relocated and rescaled to have zero mean and unit variance. Tail conditions ensure that the reconstructed density is integrable. The resulting identified innovation density and its score are used consistently in the conditional likelihood, the volatility update, and the construction of VaR and ES forecasts. The neural component therefore allows the innovation distribution and the response of volatility to new returns to be learned within a common probabilistic framework.

We consider two estimation procedures. First, the two-step GAS-NN procedure estimates the raw neural score by regularized score matching on unconditionally standardized returns. After density reconstruction and location--scale identification, the neural parameters are held fixed while the GAS parameters are estimated by conditional maximum likelihood. This first stage supplies a data-based working density, since unconditionally standardized returns still contain time-varying conditional volatility. The two-step procedure separates neural score estimation from estimation of the volatility dynamics.

Second, the one-step GAS-NN procedure jointly estimates the neural and GAS parameters using a regularized conditional likelihood, initialized from the corresponding two-step solution. Joint refinement allows the innovation density and the standardized residuals implied by the volatility filter to adjust together. The comparison between the two procedures therefore examines whether adapting the neural score to the conditional volatility model improves forecasting relative to holding the score-matching estimate fixed.

The contribution of this paper is threefold. First, we develop a neural representation of the innovation-density score within the GAS framework, preserving the connection between a valid innovation density and the volatility updating function. Second, we develop and compare two-step score-matching estimation and one-step joint likelihood refinement, providing alternative ways to connect neural density learning with volatility estimation. Third, we evaluate the proposed models using eight international equity indices, comparing their predictive density, VaR, and joint VaR--ES forecasts with Gaussian and Student-$t$ GAS benchmarks. The analysis considers both the full out-of-sample period and the first half of 2020, allowing the forecasting implications of the learned scores to be examined across markets and during the initial COVID-19 disruption.

The remainder of the paper is organized as follows. Section~\ref{background-models} reviews score-driven volatility models and the parametric benchmarks. Section~\ref{proposed-models} introduces GAS-NN, and Section~\ref{model-estimation} describes density reconstruction and the two estimation procedures. Section~\ref{sec:data_empirical_study} presents the data, architecture selection, and in-sample results. Section~\ref{out-of-sample-analysis} reports the forecasting analysis, and Section~\ref{conclusion} concludes.

\section{Background models}
\label{background-models}

\subsection{Score-driven volatility models}
\label{background-gas}

We consider the observation model
\begin{equation}
\label{eq:gas-observation}
    r_t=\exp(f_t/2)z_t,
    \qquad
    f_t=\log\sigma_t^2,
    \qquad
    z_t\overset{\mathrm{i.i.d.}}{\sim}\mathcal D(0,1),
\end{equation}
where $r_t$ is a demeaned percentage log return.
Let $C_t$ denote the closing price and define the raw percentage
return as $R_t=100\log(C_t/C_{t-1})$.
Within each estimation window, we use
$r_t=R_t-\bar R$, where $\bar R$
is the window's sample mean. This mean is held fixed when
constructing forecasts from that window.
The conditional variance is
$\sigma_t^2=\operatorname{Var}(r_t\mid\mathcal F_{t-1})$,
and $f_t$ is the conditional log variance.
Here $\mathcal F_{t-1}$ denotes the information
available before observing $r_t$, with respect to
which $f_t$ is measurable.
The notation $\mathcal D(0,1)$ denotes an innovation
distribution with mean zero and variance one,
whose density is denoted by $p$.
The innovations are independent of
$\mathcal F_{t-1}$, and their density remains fixed
within each estimation window.

Following the generalized autoregressive score
(GAS) framework of \citet{creal2013} and
\citet{harvey2013}, the log variance evolves according to
\begin{equation}
\label{eq:gas-recursion}
\begin{aligned}
   & f_{t+1}
       =\omega+\alpha s_t+\beta f_t,\\
   & s_t=S_t\nabla_t,\\
    &\nabla_t
       =\frac{\partial\log p(r_t\mid f_t,\mathcal F_{t-1};
                            \theta_{\mathrm{GAS}})}
              {\partial f_t}.
\end{aligned}
\end{equation}

Under the specified conditional density and standard
likelihood regularity and integrability conditions,
$E[s_t\mid\mathcal F_{t-1}]=0$.
For a stationary log-variance process with finite mean,
taking expectations in Equation~\eqref{eq:gas-recursion} gives
$E[f_t]=\omega/(1-\beta)$. Thus, $\omega$ determines
the long-run mean of the log variance together with
$\beta$. The coefficient $\beta$ controls the serial
dependence of $f_t$, while $\alpha$ scales the score
contribution to $f_{t+1}$.

The parameter vector
is $\theta_{\mathrm{GAS}}=(\omega,\alpha,\beta)$.
In the empirical specifications, $\omega\in\mathbb R$,
$\alpha>0$, and $0<\beta<1$; the transformations
used to impose these restrictions are given in
Appendix~\ref{app:implementation}.
In Equation~\eqref{eq:gas-recursion}, $\nabla_t$ is the
conditional log-likelihood score with respect to $f_t$,
and $S_t$ is a positive,
$\mathcal F_{t-1}$-measurable scaling factor.
Common choices for $S_t$ are given in Equation
\eqref{eq:gas-information-scaling}.

For a fixed innovation density $p$, the conditional
return density is
\begin{equation}
\label{eq:gas-return-density}
\begin{aligned}
    p(r_t\mid f_t,\mathcal F_{t-1};\theta_{\mathrm{GAS}})
      &=\left|\frac{\partial z_t}{\partial r_t}\right|p(z_t)
        \\
      &=\exp(-f_t/2)p(z_t),
\end{aligned}
\end{equation}
where $\exp(-f_t/2)$ is the absolute Jacobian
of the transformation from $r_t$ to $z_t$.
Taking logarithms therefore expresses the
conditional log-likelihood contribution of the
observed return in terms of $z_t$:
\begin{equation}
\label{eq:gas-likelihood}
    \log p(r_t\mid f_t,\mathcal F_{t-1};\theta_{\mathrm{GAS}})
      =\log p(z_t)-\frac12 f_t.
\end{equation}

Differentiating with respect to $f_t$ gives
\begin{equation}
\label{eq:gas-score-derivative}
\begin{aligned}
    \nabla_t
       &=\frac{\partial\log p(r_t\mid f_t,\mathcal F_{t-1};
                             \theta_{\mathrm{GAS}})}
               {\partial f_t}\\
       &=\frac{\partial\log p(z_t)}{\partial z_t}
         \frac{\partial z_t}{\partial f_t}
         -\frac12\\
       &=-\frac12\left\{
           1+z_t\frac{\partial\log p(z_t)}
                        {\partial z_t}
         \right\},
\end{aligned}
\end{equation}
where $\partial z_t/\partial f_t=-z_t/2$ when
the observed return and all static parameters,
including those of the innovation density, are held fixed.
The derivative $\partial\log p(z_t)/\partial z_t$
is the innovation-density score. It measures the
local slope of the log density and determines how
the assumed innovation distribution shapes the
volatility response to a new observation.

To express this dependence explicitly, define the
negative innovation-density score as
\begin{equation}
\label{eq:phi-definition}
    \phi(z)
       =-\frac{\partial\log p(z)}{\partial z}.
\end{equation}

The unscaled score $\nabla_t$ and the updating term
$s_t$ in Equation~\eqref{eq:gas-recursion} can then be written as
\begin{equation}
\label{eq:gas-score-phi}
    \nabla_t=\frac12\{z_t\phi(z_t)-1\},
    \qquad
    s_t=\frac{S_t}{2}\{z_t\phi(z_t)-1\}.
\end{equation}

Substituting the expression for $s_t$ into Equation 
\eqref{eq:gas-recursion} expresses the GAS update
in terms of the standardized innovation $z_t$
and its negative innovation-density score $\phi(z_t)$:
\begin{equation}
\label{eq:gas-recursion-phi}
    f_{t+1}
       =\omega+\frac{\alpha S_t}{2}
          \{z_t\phi(z_t)-1\}+\beta f_t.
\end{equation}

This representation makes explicit how the
innovation-density score determines the volatility
response to a new observation.

The scaling factor $S_t$ in Equation~\eqref{eq:gas-recursion}
is commonly chosen as
\begin{equation}
\label{eq:gas-information-scaling}
    S_t\in\left\{
        1,\,\mathcal I_t^{-1/2},\,\mathcal I_t^{-1}
    \right\},
\end{equation}
where the three choices correspond to unit scaling,
inverse square-root information scaling, and inverse
information scaling, respectively \citep{creal2013}.
The quantity $\mathcal I_t$ denotes the conditional
Fisher information for $f_t$.

An advantage of the log-variance parameterization
$f_t=\log\sigma_t^2$ in Equation~\eqref{eq:gas-observation} is that the unscaled score $\nabla_t$
can be expressed solely as a function of the
standardized innovation $z_t$. For a fixed innovation
density, its conditional Fisher information therefore
does not depend on the time-varying parameter $f_t$.
Under standard likelihood regularity conditions,
\begin{equation}
\label{eq:gas-information-moment}
    \mathcal I_t
      =E_{t-1}[\nabla_t^2]
      =\frac14 E_{t-1}
         \left[\{z_t\phi(z_t)-1\}^2\right],
\end{equation}
where $\mathcal I_t$ is assumed finite and strictly
positive.

The information-based choices for $S_t$ therefore
multiply the score by a positive constant.
Since the scaled score $s_t$ is multiplied by $\alpha$
in Equation~\eqref{eq:gas-recursion}, this constant
can be absorbed into the freely estimated coefficient
$\alpha$. We consequently use unit scaling,
$S_t=1$, throughout, so that $s_t=\nabla_t$.

\subsection{Parametric GAS specifications}
\label{background-parametric-gas}

Parametric specifications such as the Gaussian and
Student-$t$ models prescribe the functional form of
the negative innovation-density score $\phi$ in
Equation~\eqref{eq:phi-definition}. Under unit scaling,
these forms determine the updating term $s_t$ in
Equation~\eqref{eq:gas-score-phi} and the volatility
dynamics in Equation~\eqref{eq:gas-recursion-phi}.

For standard Gaussian innovations, these terms are
\begin{equation}
\label{eq:gaussian-gas-scores}
\begin{aligned}
   & \phi(z)=z,\\
   & s_t=\frac12(z_t^2-1).
\end{aligned}
\end{equation}

Substituting the updating term in
Equation~\eqref{eq:gaussian-gas-scores} into
Equation~\eqref{eq:gas-recursion} gives the Gaussian
GAS model (GAS-N). Its log-variance score grows
quadratically with the magnitude of the standardized
innovation.

For Student-$t$ innovations standardized to have
unit variance, the corresponding terms are
\begin{equation}
\label{eq:student-t-gas-scores}
\begin{aligned}
    &\phi(z)
       =\frac{(\nu+1)z}{\nu-2+z^2},\\
   & s_t
       =\frac12\left\{
           \frac{(\nu+1)z_t^2}{\nu-2+z_t^2}-1
         \right\},
    \qquad \nu>2,
\end{aligned}
\end{equation}
where $\nu$ is the degrees-of-freedom parameter,
with $\nu>2$ ensuring that the innovation variance
exists. Substituting the updating term in
Equation~\eqref{eq:student-t-gas-scores} into
Equation~\eqref{eq:gas-recursion} gives the Student-$t$
GAS model (GAS-t). Its log-variance score approaches
$\nu/2$ as $|z_t|\to\infty$, bounding the immediate
score contribution of extreme observations.

Both specifications use the same log-variance
recursion in Equation~\eqref{eq:gas-recursion}.
Their log-variance scores are even functions of
$z_t$, so standardized innovations of equal
magnitude and opposite signs produce the same
score contribution to the volatility update.
The different negative innovation-density score
shapes are illustrated in
Figure~\ref{fig:ftse-negative-scores}:
the Gaussian score is linear, whereas the Student-$t$
score approaches zero as $|z|\to\infty$.

\section{Proposed models}
\label{proposed-models}

Parametric innovation distributions prescribe the form of the density score.
We instead estimate a flexible negative density score with a feedforward
neural network. Density reconstruction and location--scale identification
connect this representation to a valid innovation density and its
corresponding volatility update. The neural representation can
accommodate asymmetric density shapes, including negative skewness.

The identified neural negative innovation-density score is defined as
\begin{equation}
\label{eq:neural-phi}
    \mathrm{NN}(z)=\phi(z)
       =-\frac{\partial\log p(z)}{\partial z},
\end{equation}
where $p$ denotes the density of the standardized
innovations, with mean zero and unit variance.

Replacing $\phi$ in Equation~\eqref{eq:gas-recursion-phi}
with its neural representation $\mathrm{NN}$ and using
unit scaling, $S_t=1$, gives the proposed GAS-NN model:
\begin{equation}
\label{eq:gas-nn-model}
\begin{aligned}
    &r_t=\exp(f_t/2)z_t,
    \qquad z_t\overset{\mathrm{i.i.d.}}{\sim}\mathcal D(0,1),\\
    &f_{t+1}
       =\omega+\frac{\alpha}{2}
          \left\{z_t\mathrm{NN}(z_t)-1\right\}+\beta f_t,
\end{aligned}
\end{equation}
where $\mathrm{NN}(z_t)$ represents the estimated
negative innovation-density score evaluated at the
standardized innovation $z_t$.

The neural network selected for the empirical analysis has two hidden layers
with eight neurons in each layer, one scalar input,
and one scalar output, giving a $1$--$8$--$8$--$1$
architecture. The architecture selection procedure
is described in Section~\ref{sec:architecture_selection}.
The neural network can be written as
\begin{equation}
\label{eq:raw-neural-network}
\begin{aligned}
    &\mathbf h^{(1)}
       =\tanh\!\left(\mathbf W^{(1)}u
                       +\mathbf b^{(1)}\right),\\
    &\mathbf h^{(2)}
       =\tanh\!\left(\mathbf W^{(2)}\mathbf h^{(1)}
                       +\mathbf b^{(2)}\right),\\
    &g(u)
       =\mathbf W^{(3)}\mathbf h^{(2)}+b^{(3)},
\end{aligned}
\end{equation}
where $\mathbf h^{(1)}$ and $\mathbf h^{(2)}$
are the hidden-layer activation vectors, and $\tanh$
acts elementwise. The matrices $\mathbf W^{(1)}$ and
$\mathbf W^{(2)}$ contain the hidden-layer weights,
and $\mathbf b^{(1)}$ and $\mathbf b^{(2)}$ are the
corresponding bias vectors. The output weights
$\mathbf W^{(3)}$ form a row vector, and $b^{(3)}$ is
a scalar bias. The function $g$ denotes the raw neural negative
density score, and $u$ is a generic scalar input. The inputs used
in the two estimation procedures are described in
Sections~\ref{sec:estimation-score-matching} and~\ref{proposed-one-step}.
The smooth $\tanh$ activations make $g$ differentiable with respect
to its input, as required for score matching; the output layer is linear.
The vector $\theta_{\mathrm{NN}}$ collects all network weights and biases. 

The identified neural score $\mathrm{NN}(z)$ is obtained by relocating and rescaling the raw score $g$, given by
\begin{equation}
\label{eq:identified-neural-score}
    \mathrm{NN}(z)=a\,g(m+az),
\end{equation}
where $m$ and $a>0$ denote the mean and standard
deviation, respectively, of the raw density $q(u)$
reconstructed from $g$. We relocate and rescale this
raw density to obtain the identified innovation density
$p(z)$ with mean zero and unit variance, as explained
in Section~\ref{estimation-identification}.

We consider two procedures for estimating the model.
Two-step GAS-NN estimation, described in
Section~\ref{proposed-two-step}, first learns the raw
neural score by score matching, constructs its
identified density and score, and then holds these
functions fixed while estimating $\theta_{\mathrm{GAS}}$
by conditional maximum likelihood. One-step GAS-NN
estimation, described in Section~\ref{proposed-one-step},
jointly updates $\theta_{\mathrm{NN}}$ and
$\theta_{\mathrm{GAS}}$ using the regularized conditional
likelihood, initialized at the two-step estimates.
The estimation objectives and procedures are detailed
in Section~\ref{model-estimation}.

\section{Model estimation}
\label{model-estimation}

\subsection{Density reconstruction and location--scale identification}
\label{estimation-identification}

The innovation density must satisfy
\begin{equation}
\label{eq:innovation-identification-conditions}
    \int_{\mathbb R}p(z)\,dz=1,
    \qquad
    E(z_t)=0,
    \qquad
    \operatorname{Var}(z_t)=1,
\end{equation}
where $p$ is the innovation density, and the
expectation and variance are evaluated under this
density. The first condition normalizes its total
probability mass. The remaining conditions ensure
that the observation equation in
\eqref{eq:gas-observation} has conditional mean
zero and conditional variance $\exp(f_t)$,
identifying $f_t$ as the conditional log variance.

The network in Equation~\eqref{eq:raw-neural-network} represents a raw negative density score.
Recovering a density from this score requires
normalization, and this reconstructed raw density need
not have zero mean or unit variance. We therefore
normalize, relocate, and rescale the raw density to
satisfy Equation~\eqref{eq:innovation-identification-conditions}.
Such location--scale standardization is also used
by \citet{blasques2016}.

The raw density reconstruction is given by
\begin{equation}
\label{eq:raw-neural-density}
\begin{aligned}
    &H(u)=\int_0^u g(v)\,dv,\\
    &q(u)=\frac{\exp\{-H(u)\}}{C},\\
    &C=\int_{\mathbb R}\exp\{-H(v)\}\,dv,
\end{aligned}
\end{equation}
where $g$ is the raw neural negative density
score, $H$ is its integral, $q$ is the
reconstructed raw density before location--scale
standardization, and $C$ is the normalization
constant, which must be finite. Choosing zero as the lower
integration limit fixes $H(0)=0$ without restricting the density:
an additive constant in $H$ cancels between the numerator and $C$.

The mean $m$ and variance $a^2$ of the
reconstructed raw density $q$ are given by
\begin{equation}
\label{eq:raw-neural-moments}
    m
       =\int_{\mathbb R}u q(u)\,du,
    \qquad
    a^2
       =\int_{\mathbb R}(u-m)^2q(u)\,du,
\end{equation}
where $a>0$ is the standard deviation of $q$.
Using the standardized coordinate $z=(u-m)/a$,
we relocate and rescale the raw density $q$
and its negative density score $g$ to obtain
the identified innovation density $p$ with
mean zero and unit variance and its corresponding
negative density score $\mathrm{NN}$:
\begin{equation}
\label{eq:identified-neural-density}
\begin{aligned}
   & p(z)
       =a q(m+az),\\
    &\mathrm{NN}(z)=\phi(z)
       =a g(m+az),
\end{aligned}
\end{equation}

To ensure that the normalization constant $C$ is finite, the neural
score is evaluated on a reconstruction interval
$[-B,B]$, where $B>0$ determines the
reconstruction boundary. Outside this interval,
the score is held constant at its corresponding
endpoint value: $g(u)=g(-B)$ for $u<-B$
and $g(u)=g(B)$ for $u>B$.
Whenever $g$ and $H$ are used for density
reconstruction or to evaluate the identified
density and score, they include this continuation.
The endpoint sign conditions $g(-B)<0<g(B)$
ensure that $H(u)$ increases linearly toward
$+\infty$ in both tails. Consequently,
$\exp\{-H(u)\}$ decays exponentially,
ensuring a finite normalization constant $C$
and finite moments of $q$.
The boundary penalty introduced in
Section~\ref{estimation-regularization}
encourages these required endpoint signs.
All normalization and moment calculations
include the continued tails.

Dividing by $C$ gives a valid density $q$ with unit total mass.
The location--scale transformation imposes zero
mean and unit variance while preserving this
normalization:
\begin{equation}
\label{eq:normalization-preservation}
    \int_{\mathbb R}p(z)\,dz
      =\int_{\mathbb R}
         a q(m+az)\,dz
      =\int_{\mathbb R}q(u)\,du
      =1,
\end{equation}
where the change of variable is $u=m+az$.
Thus, density normalization and location--scale
transformation together satisfy all three conditions
in Equation~\eqref{eq:innovation-identification-conditions}.
The GAS recursion uses the identified score
$\mathrm{NN}$, evaluated at the conditionally
standardized residual $z_t$.

The quantities $C$, $m$, and $a$
are deterministic functions of the neural parameters
$\theta_{\mathrm{NN}}$. For two-step GAS-NN, they are calculated
after score matching and remain
fixed during the subsequent likelihood estimation
of the GAS parameters $\theta_{\mathrm{GAS}}$. For one-step GAS-NN,
they are recalculated whenever $\theta_{\mathrm{NN}}$ changes,
and gradients propagate through the normalization
and moment calculations. They are therefore updated
through $\theta_{\mathrm{NN}}$ during joint estimation and are
not additional free parameters.

The identified innovation density and its negative
density score satisfy the following identities under
suitable differentiability and integrability conditions,
with $p(z)$ and $zp(z)$ vanishing in both tails:
\begin{equation}
\label{eq:innovation-score-identities}
    E[\phi(z_t)]=0,
    \qquad
    E[z_t\phi(z_t)]=1,
\end{equation}
where expectations are taken under the identified
density $p$, and $\phi=\mathrm{NN}$ is the identified
neural score in Equation~\eqref{eq:identified-neural-density}.
These identities follow from integration by parts.
Under correct specification, $z_t$ is independent of
$\mathcal F_{t-1}$, so the identities also hold
conditionally on past information. The first gives
$E[\phi(z_t)\mid\mathcal F_{t-1}]=0$, while the
second gives
\begin{equation}
\label{eq:zero-mean-volatility-score}
    E[s_t\mid\mathcal F_{t-1}]
      =\frac12\left\{
        E[z_t\phi(z_t)\mid\mathcal F_{t-1}]-1
        \right\}
      =0,
\end{equation}
where $s_t$ is the unit-scaled log-variance score
entering the GAS recursion.
This establishes the conditional mean-zero score
property introduced after Equation~\eqref{eq:gas-recursion}.

\subsection{Score-matching estimation}
\label{sec:estimation-score-matching}

Score matching estimates the raw neural score in the
first stage of two-step GAS-NN. The resulting estimates
also provide the initialization for one-step GAS-NN.

The first-stage training inputs are constructed as
\begin{equation}
\label{eq:score-matching-inputs}
    u_t=\frac{r_t}
      {\sqrt{\left(\frac{1}{T_{\mathrm{in}}}
             \sum_{j=1}^{T_{\mathrm{in}}}r_j^2\right)}},
\end{equation}
where $u_t$ is the standardized input to the raw neural
network and $T_{\mathrm{in}}$ is the number of observations
in the estimation sample. The denominator is the sample
standard deviation of the demeaned returns, calculated
with divisor $T_{\mathrm{in}}$. Dividing by this common
scale gives inputs with sample mean zero and sample
second moment one.

The inputs $u_t$ differ from the conditional standardized
residuals $z_t=r_t\exp(-f_t/2)$ used in the GAS recursion.
The former use a single sample-wide scale, while the
latter use the time-varying conditional standard
deviation. Consequently, score matching on $u_t$ learns
the shape of the marginal standardized-return
distribution, which generally differs from the
conditional innovation distribution. The fitted density
is subsequently normalized and standardized for use
as the GAS innovation density.

The target negative density score is
\begin{equation}
\label{eq:input-negative-score}
    g^{*}(u)=-\frac{\partial\log p_U(u)}{\partial u},
\end{equation}
where $p_U(u)$ is the unknown marginal density of the
standardized training inputs. The raw neural score
$g(u)$ approximates this target. 

Following \citet{hyvarinen2005}, the population
score-matching criterion is
\begin{equation}
\label{eq:population-score-matching}
    J_{\mathrm{pop}}(\theta_{\mathrm{NN}})=\frac12\int_{\mathbb R}
       \left\{g(u)-g^{*}(u)\right\}^2
       p_U(u)\,du,
\end{equation}
where $g(u)$ is the neural approximation governed by
$\theta_{\mathrm{NN}}$. The criterion measures one-half
of the expected squared difference between the neural
score and the unknown target score.

Under suitable differentiability, integrability, and
boundary conditions, integration by parts removes
the unknown target score from the parameter-dependent
part of the criterion. Omitting an additive term
independent of $\theta_{\mathrm{NN}}$ and replacing the
expectation with a sample average gives
\begin{equation}
\label{eq:empirical-score-matching}
    J(\theta_{\mathrm{NN}})
      =\frac1{T_{\mathrm{in}}}\sum_{t=1}^{T_{\mathrm{in}}}
        \left\{\frac12g(u_t)^2-g'(u_t)\right\},
\end{equation}
where $J$ is the empirical score-matching objective and
$g'(u)=\partial g(u)/\partial u$ is the derivative of
the network output with respect to its scalar input.
This objective is evaluated directly from the training
inputs and the network, without estimating $p_U(u)$ or
computing the normalization constant of the reconstructed
density.

Training adds the boundary and neural parameter penalties
from Section~\ref{estimation-regularization} to
$J(\theta_{\mathrm{NN}})$. The resulting neural parameter
estimate, $\widehat\theta_{\mathrm{NN}}$, determines
the identified density $p$ and score $\phi$ through
Equations~\eqref{eq:raw-neural-density}--\eqref{eq:identified-neural-density}.
These functions remain fixed during the second-stage
likelihood estimation of the GAS parameters.

\subsection{Conditional maximum likelihood}
\label{estimation-likelihood}

Let $\boldsymbol{\theta}=(\theta_{\mathrm{GAS}},
\theta_{\mathrm{NN}})$ denote the complete GAS-NN
parameter vector. The mean negative conditional log
likelihood is
\begin{equation}
\label{eq:neural-conditional-nll}
\begin{aligned}
    \mathcal L_{\mathrm{NLL}}(\boldsymbol{\theta})
      &=-\frac1{T_{\mathrm{in}}}\sum_{t=1}^{T_{\mathrm{in}}}
        \log p(r_t\mid f_t,\mathcal F_{t-1};\boldsymbol{\theta})\\
      &=\frac1{T_{\mathrm{in}}}\sum_{t=1}^{T_{\mathrm{in}}}
        \left\{\frac12 f_t-\log p(z_t)\right\},
\end{aligned}
\end{equation}
where $f_t$ is the conditional log variance generated
by the GAS-NN recursion, initialized at
$f_1=\log\{T_{\mathrm{in}}^{-1}\sum_{t=1}^{T_{\mathrm{in}}}r_t^2\}$.
The quantity $z_t=r_t\exp(-f_t/2)$ is the corresponding conditional
standardized residual, and $p(z)$ is the identified
innovation density. Both $f_t$ and $z_t$ depend on
$\boldsymbol{\theta}$ through the recursion, while
$p(z)$ depends on $\theta_{\mathrm{NN}}$ through density
reconstruction and identification. These dependencies
are suppressed in their notation. The transformation
in Equation~\eqref{eq:gas-return-density} therefore continues
to apply, with the full parameter vector
$\boldsymbol{\theta}$ replacing $\theta_{\mathrm{GAS}}$
when the innovation density is estimated jointly.

The likelihood is evaluated using demeaned percentage
returns $r_t$. Division by the sample standard deviation
is used only to construct the first-stage score-matching
inputs $u_t$ in Equation~\eqref{eq:score-matching-inputs}.

The Gaussian and Student-$t$ benchmarks are estimated
by the same conditional likelihood principle using
their respective innovation densities. For GAS-t,
the degrees-of-freedom parameter $\nu$ is estimated
jointly with $\theta_{\mathrm{GAS}}$, subject to $\nu>2$
so that the innovation variance exists.

\subsection{Regularization and complexity control}
\label{estimation-regularization}

The boundary penalty is
\begin{equation}
\label{eq:neural-boundary-penalty}
    P_B
      =[\tau+g(-B)]_+^2
       +[\tau-g(B)]_+^2,
\end{equation}
where $g$ is the raw neural score, $B>0$
is the reconstruction boundary, $\tau>0$ is a
prescribed margin, and $[x]_+=\max(x,0)$ denotes
the positive part. The penalty is zero when
$g(-B)\leq-\tau$ and
$g(B)\geq\tau$.

These inequalities give a negative left-tail slope
and a positive right-tail slope for $H$.
Under the constant-score continuation, $H$
therefore increases toward $+\infty$ in both tails,
so the unnormalized density decays exponentially
and $C$ is finite. Normalization is then
imposed explicitly by dividing by $C$,
as described in
Section~\ref{estimation-identification}.

Valid density reconstruction requires
the strict sign conditions
$g(-B)<0<g(B)$.

To control neural network complexity, we also
apply an $L_2$ penalty to the network weights
and biases, given by
$\|\theta_{\mathrm{NN}}\|_2^2$.
The coefficients $\lambda_B,\lambda_W\geq0$
control the boundary and neural parameter
penalties, respectively.
The latter corresponds to the coupled neural
weight decay used during optimization.
Regularization is applied only to the neural
parameters, while the GAS parameters
$\theta_{\mathrm{GAS}}$ remain unpenalized.
Both penalties are incorporated into the
estimation objectives specified in Equation~\eqref{eq:two-step-procedure} and Equation~\eqref{eq:one-step-procedure}.

\subsection{Two-step GAS-NN estimation}
\label{proposed-two-step}

The two-step procedure estimates the neural score
and the volatility dynamics sequentially:
\begin{subequations}
\label{eq:two-step-procedure}
\begin{align}
    &\widehat\theta_{\mathrm{NN}}
       =\operatorname*{arg\,min}_{\theta_{\mathrm{NN}}}
         \left\{J(\theta_{\mathrm{NN}})+\lambda_B P_B
         +\lambda_W\|\theta_{\mathrm{NN}}\|_2^2\right\},\\
    &\widehat\theta_{\mathrm{GAS}}
       =\operatorname*{arg\,min}_{\theta_{\mathrm{GAS}}}
         \mathcal L_{\mathrm{NLL}}
         (\theta_{\mathrm{GAS}},\widehat\theta_{\mathrm{NN}}),
         \label{eq:two-step-mle}
\end{align}
\end{subequations}
where $\widehat\theta_{\mathrm{NN}}$ is the regularized
score-matching estimate of the neural weights and
biases, and $\widehat\theta_{\mathrm{GAS}}$ is the
subsequent conditional maximum likelihood estimate
of the GAS parameters. The GAS parameters obey the restrictions
in Section~\ref{background-gas}. Retained neural estimates must
satisfy the strict endpoint sign conditions before they define
a valid reconstructed density for likelihood evaluation.
The boundary penalty encourages these signs during training;
it does not impose a hard constraint on every intermediate iterate.
The empirical score-matching
loss $J$ is defined in Equation~\eqref{eq:empirical-score-matching},
the regularization terms in
Section~\ref{estimation-regularization},
and the mean negative conditional log likelihood
$\mathcal L_{\mathrm{NLL}}$ in Equation~\eqref{eq:neural-conditional-nll}.

In the first stage, the raw neural score $g$ is trained
using the standardized inputs $u_t$ from
Equation~\eqref{eq:score-matching-inputs}. Evaluating the neural
functions at $\widehat\theta_{\mathrm{NN}}$ then determines
the identified innovation density $p$ and score
$\mathrm{NN}=\phi$ through the reconstruction and
standardization in Section~\ref{estimation-identification}.

In the second stage, these functions remain fixed
while $\theta_{\mathrm{GAS}}$ is estimated by conditional
maximum likelihood. The conditional log variances
$f_t$ and standardized residuals $z_t$ continue to
change with $\theta_{\mathrm{GAS}}$ through the GAS
recursion. Thus, the second stage evaluates the
fixed neural score at residuals generated by the
candidate volatility parameters.

\subsection{One-step GAS-NN estimation}
\label{proposed-one-step}

The one-step procedure jointly estimates the GAS
and neural parameters:
\begin{equation}
\label{eq:one-step-procedure}
\begin{aligned}
    (\widehat\theta_{\mathrm{GAS}},\widehat\theta_{\mathrm{NN}})
       &=\operatorname*{arg\,min}_{\theta_{\mathrm{GAS}},\theta_{\mathrm{NN}}}
         \Bigl\{\mathcal L_{\mathrm{NLL}}(\boldsymbol{\theta})
         +\lambda_B P_B
         +\lambda_W\|\theta_{\mathrm{NN}}\|_2^2\Bigr\},
\end{aligned}
\end{equation}
where $\widehat\theta_{\mathrm{GAS}}$ and
$\widehat\theta_{\mathrm{NN}}$ are the joint estimates
of the GAS parameters and neural weights and biases,
respectively. The objective combines the mean negative
conditional log likelihood in Equation~\eqref{eq:neural-conditional-nll}
with the boundary and neural parameter penalties
described in Section~\ref{estimation-regularization}.
Optimization is initialized at the two-step estimates
from Section~\ref{proposed-two-step}. The same hat
notation denotes the estimates under the procedure
being described. The GAS restrictions and acceptance conditions
for the retained neural density are the same as in two-step estimation.

During joint estimation, changes in $\theta_{\mathrm{NN}}$
update the identified innovation density $p$ and score
$\mathrm{NN}$, including the normalization and
location--scale quantities $C$, $m$, and $a$.
Changes in either parameter group can also alter
the conditional log variances $f_t$ and standardized
residuals $z_t$. Gradients therefore propagate through
density reconstruction, location--scale identification,
and the complete volatility recursion.

Both procedures use the common GAS-NN specification
in Equation~\eqref{eq:gas-nn-model}. Two-step estimation
holds the neural parameters fixed during likelihood
estimation, whereas one-step estimation updates
them jointly with the GAS parameters. This allows
the innovation density and volatility dynamics
to adapt together through the conditional likelihood.
The argmin expressions specify the estimation
targets; numerical estimates are obtained using
the prescribed optimization settings and training
budgets.

\section{Data and empirical study}
\label{sec:data_empirical_study}

\subsection{Data description}
\label{sec:data_description}

The empirical analysis considers eight equity indices: the S\&P 500,
FTSE, GDAXI, Australian All Ordinaries (AORD), Dow Jones (DJI), Hang Seng (HSI),
Nikkei 225 (N225), and EURO STOXX 50 (STOXX50E). The sample spans January 2004 to May 2026,
with 5,483--5,743 daily observations per market. Each market series follows its own trading
calendar, excluding market-specific non-trading days and observations
with missing closing prices. Table~\ref{tab:samples} reports descriptive
statistics for the raw percentage returns and the total sample sizes.
Returns are demeaned within each estimation window before fitting the models.

The sample is divided approximately 70:30 into in-sample and
out-of-sample periods. The in-sample period extends from January 2004
to December 2019 and includes the 2008 Global Financial Crisis. The
out-of-sample forecasting period runs from January 2020 to 29 May 2026,
covering the onset of the COVID-19 pandemic and the subsequent market
disruption. We also evaluate the first half of 2020, from each market's first
available January observation to 30 June 2020, as a COVID-period
subsample. This window includes the World Health Organization's
declaration of a public health emergency of international concern
on 30 January and its characterization of the outbreak as a pandemic
on 11 March \citep{who2026}.

\begin{table}[H]
    \centering
    \small
    \caption{Descriptive statistics for the return series of eight equity indices}
    \label{tab:samples}
    \setlength{\tabcolsep}{3pt}
    \renewcommand{\arraystretch}{1.15}
    \begin{tabular*}{\linewidth}{@{\extracolsep{\fill}}lrrrrrrr@{}}
    \toprule
    \textbf{Market} & \textbf{N} & \textbf{Mean}
    & \textbf{Std. dev.} & \textbf{Min} & \textbf{Max}
    & \textbf{Skewness} & \shortstack{\textbf{Excess}\\\textbf{kurtosis}} \\
    \midrule
    S\&P 500
    & 5,632 & 0.0341 & 1.1850 & -12.7652 & 10.9572 & -0.4785 & 13.4934 \\
    FTSE
    & 5,661 & 0.0149 & 1.0760 & -11.5124 & 9.3843 & -0.4449 & 10.8822 \\
    GDAXI
    & 5,694 & 0.0324 & 1.2823 & -13.0549 & 10.7975 & -0.2287 & 8.2964 \\
    AORD
    & 5,671 & 0.0176 & 0.9952 & -10.0094 & 6.3548 & -0.7841 & 8.4020 \\
    DJI
    & 5,637 & 0.0281 & 1.1191 & -13.8418 & 10.7643 & -0.4322 & 16.3890 \\
    HSI
    & 5,520 & 0.0126 & 1.4693 & -14.1833 & 13.4068 & -0.1329 & 8.7799 \\
    N225
    & 5,483 & 0.0333 & 1.4470 & -13.2341 & 13.2346 & -0.4577 & 8.5502 \\
    STOXX50E
    & 5,743 & 0.0137 & 1.2980 & -13.2405 & 10.4376 & -0.3132 & 8.3008 \\
    \bottomrule
    \end{tabular*}

    \par\vspace{6pt}
    \begin{minipage}{\linewidth}
    \footnotesize\raggedright
    \textit{Notes:} Statistics are calculated using percentage log
    returns from January 2004 to 29 May 2026. $N$ denotes the number
    of daily observations. Skewness and excess kurtosis are bias-corrected
    sample estimates. Excess kurtosis is zero for a Gaussian
    distribution.
    \end{minipage}
\end{table}

\subsection{Neural network architecture selection and final estimation}
\label{sec:architecture_selection}

The two-step and one-step GAS-NN models are denoted GAS-NN-2
and GAS-NN-1, respectively. A common neural network architecture
is selected using the S\&P 500. Its in-sample period is divided
approximately 75:25 into a training sample of 3,016 observations,
from January 2004 to December 2015, and a validation sample of
1,006 observations, from January 2016 to December 2019.
Predictive validation likelihoods are computed by continuing
the fitted GAS filter through the validation period, with
model parameters held at their training-sample estimates.

We consider five feedforward architectures: 1--8--1,
1--16--1, 1--8--8--1, 1--16--16--1, and 1--32--32--1,
where the numbers indicate the neurons in successive input,
hidden, and output layers. Architecture selection uses
boundary margin $\tau=0.01$. For GAS-NN-2, validation is performed
every 10 score-matching epochs. At each checkpoint, the neural
density is reconstructed and identified to have mean zero and
unit variance, the GAS parameters are estimated on the training
sample, and the resulting model is evaluated using the validation
predictive mean negative log-likelihood (NLL). GAS-NN-1 is
initialized from the best GAS-NN-2 solution for the same
architecture and evaluated every 10 joint updates.

For each procedure, training stops after 10 valid validation
checks without a relative improvement exceeding $10^{-4}$,
subject to a maximum of 1,000 neural training updates.
Invalid checkpoints do not count toward the stopping rule.
The checkpoint with the lowest valid validation NLL is retained,
including improvements smaller than the patience threshold.

A common architecture is selected by minimizing the equally
weighted criterion
\begin{equation}
\label{eq:architecture_selection}
    V(A)=\frac{V_2(A)+V_1(A)}{2},
\end{equation}
where $A$ denotes a candidate neural network architecture,
and $V_2(A)$ and $V_1(A)$ denote its lowest validation
predictive mean NLL under GAS-NN-2 and GAS-NN-1 estimation,
respectively. Architectures whose shared scores lie within
an absolute $10^{-4}$ of the minimum are treated as
practically tied, with preference given to the architecture
with fewer neural network parameters.

Table~\ref{tab:architecture} reports the selection results.
The 1--8--8--1 network, with two hidden layers of eight
neurons and 97 trainable neural parameters, achieves the
lowest GAS-NN-2 validation NLL of 1.0081 and a GAS-NN-1
validation NLL of 1.0073. It attains the lowest shared
score of 1.0077. Its best validation checkpoints occur
after 30 score-matching epochs for GAS-NN-2 and 20
additional joint updates for GAS-NN-1. We therefore
use this architecture in all eight markets.

\begin{table}[!htbp]
    \centering
    \caption{S\&P 500 architecture selection by predictive validation NLL}
    \label{tab:architecture}
    \footnotesize
    \setlength{\tabcolsep}{4pt}
    \renewcommand{\arraystretch}{1.2}

    \begin{tabular*}{\linewidth}
        {@{\extracolsep{\fill}}lrcrrr@{}}
    \toprule
    Architecture
    & \shortstack{NN\\parameters}
    & \shortstack{Best epochs\\GAS-NN-2 / GAS-NN-1}
    & \shortstack{GAS-NN-2\\Mean NLL}
    & \shortstack{GAS-NN-1\\Mean NLL}
    & \shortstack{Shared\\score} \\
    \midrule
    1--8--1
    & 25
    & 10 / 10
    & 1.0085
    & 1.0073
    & 1.0079 \\

    1--16--1
    & 49
    & 10 / 10
    & 1.0096
    & 1.0090
    & 1.0093 \\

    \textcolor{blue}{1--8--8--1}
    & \textcolor{blue}{97}
    & \textcolor{blue}{30 / 20}
    & \textcolor{blue}{1.0081}
    & \textcolor{blue}{1.0073}
    & \textcolor{blue}{1.0077} \\

    1--16--16--1
    & 321
    & 10 / 10
    & 1.0082
    & 1.0084
    & 1.0083 \\

    1--32--32--1
    & 1,153
    & 10 / 10
    & 1.0097
    & 1.0100
    & 1.0098 \\
    \bottomrule
    \end{tabular*}

    \par\vspace{6pt}
    \begin{minipage}{\linewidth}
    \footnotesize\raggedright
    \textit{Notes:}
    GAS-NN-2 and GAS-NN-1 denote two-step and one-step
    estimation, respectively. NLL is the mean negative
    predictive log-likelihood over the validation sample.
    Best epochs identify the retained validation checkpoints,
    not the epochs at which training terminates.
    The GAS-NN-1 epoch count refers to additional joint
    updates after initialization at the retained GAS-NN-2
    solution. The shared score is the equally weighted
    average of the two unrounded validation NLL values.
    Blue identifies the selected architecture.
    Architecture selection uses boundary margin $\tau=0.01$.
    \end{minipage}
\end{table}
\FloatBarrier

For final in-sample estimation, the S\&P 500 training and
validation samples are combined, and each market is fitted
using its full in-sample period. Guided by the retained
validation checkpoints and the larger combined
training and validation sample, we adopt
30 score-matching epochs for the initial GAS-NN-2
specification and 30 additional joint updates for the
corresponding GAS-NN-1 specification. We also consider an
extended score-matching budget of 200 epochs to assess
the sensitivity of density estimation and forecasting
performance to additional first-stage training.
These final budgets are fixed experimental choices rather
than market-specific validation-selected horizons.
% All final fits use $\tau=0.01$ and the numerical settings
% reported in Appendix~\ref{app:implementation}.

Within each market, the 30- and 200-epoch neural estimates
are retained from a single 200-epoch score-matching
trajectory. The GAS parameters are estimated separately
for each retained neural density, yielding
GAS-NN-2~(30) and GAS-NN-2~(200).
Each resulting two-step solution then initializes a
corresponding one-step fit with 30 additional joint updates,
yielding GAS-NN-1~(30-30) and GAS-NN-1~(200-30).
For GAS-NN-2, the number in parentheses denotes the
score-matching budget. For GAS-NN-1, the first number
denotes the score-matching budget used to construct the
two-step initializer, and the second denotes the number
of subsequent joint updates. Holding the joint budget
fixed allows us to examine how the extent of first-stage
training affects subsequent joint estimation.
GAS-N and GAS-t denote the Gaussian and standardized
Student-$t$ benchmarks, respectively.

The selected architecture and training budgets are common
across markets, whereas return-standardization quantities,
neural weights and biases, identified innovation densities,
and GAS parameters are estimated separately for each market.

\subsection{In-sample parameter estimates and likelihood}
\label{ftse-parameter-estimates-and-training-likelihood}

We use the FTSE to illustrate the fitted models.
Its in-sample period runs from 2 January 2004 to
31 December 2019 and contains $T_{\mathrm{in}}=4{,}044$
observations. Table~\ref{tab:ftse} reports
the estimated GAS parameters, the Student-$t$ degrees
of freedom, and the mean in-sample NLL.

\begin{table}[!htbp]
    \centering
    \caption{FTSE parameter estimates and in-sample likelihood}
    \label{tab:ftse}
    \footnotesize
    \setlength{\tabcolsep}{4pt}
    \renewcommand{\arraystretch}{1.25}

    \begin{tabular*}{\linewidth}
        {@{\extracolsep{\fill}}lrrrrr@{}}
    \toprule
    Model
    & $\omega$
    & $\alpha$
    & $\beta$
    & $\nu$
    & \shortstack{Mean\\NLL} \\
    \midrule
    GAS-N
    & -0.0039
    & 0.1371
    & 0.9832
    & --
    & 1.3015 \\

    GAS-t
    & -0.0041
    & 0.2306
    & 0.9818
    & 8.2319
    & 1.2869 \\

    GAS-NN-2 (30)
    & -0.0018
    & 0.3146
    & 0.9817
    & --
    & 1.2999 \\

    GAS-NN-2 (200)
    & -0.0032
    & 0.2492
    & 0.9825
    & --
    & 1.2868 \\

    GAS-NN-1 (30-30)
    & -0.0024
    & 0.3064
    & 0.9812
    & --
    & 1.2906 \\

    GAS-NN-1 (200-30)
    & -0.0035
    & 0.2431
    & 0.9823
    & --
    & \textcolor{blue}{1.2858} \\
    \bottomrule
    \end{tabular*}

    \par\vspace{6pt}
    \begin{minipage}{\linewidth}
    \footnotesize\raggedright
    \textit{Notes:}
    All models are estimated using the same
    $T_{\mathrm{in}}=4{,}044$ observations.
    NLL is the mean conditional negative log-likelihood,
    excluding boundary and neural parameter penalties;
    lower values indicate better in-sample fit.
    For GAS-NN-2, parentheses report score-matching epochs.
    For GAS-NN-1, parentheses report score-matching epochs
    used for initialization followed by additional joint
    updates. Blue identifies the lowest mean NLL.
    \end{minipage}
\end{table}
\FloatBarrier

All six fits exhibit high log-variance persistence,
with $\beta$ ranging from approximately 0.9812 to
0.9832, indicating persistent conditional volatility
dynamics across both parametric and neural specifications.
The estimated score-response coefficient $\alpha$ is
larger for the GAS-NN models than for the parametric
GAS-N and GAS-t benchmarks. This suggests that the
neural specifications place greater weight on their
respective score innovations in updating conditional
log variance. However, the magnitude of the resulting
volatility response also depends on the scale and shape
of the corresponding score function $s_t$.
For GAS-t, the estimated degrees of freedom are 8.2319,
indicating heavier innovation tails than under the
Gaussian specification.

Replacing Gaussian innovations with standardized
Student-$t$ innovations reduces the mean in-sample
NLL from 1.3015 to 1.2869, indicating a substantial
improvement in likelihood fit from allowing heavier
innovation tails. For GAS-NN-2, extending score matching
from 30 to 200 epochs reduces the mean NLL from
1.2999 to 1.2868. The longer training budget therefore
improves the fitted neural density and brings the
two-step model's likelihood slightly below that of
GAS-t.

Joint estimation further improves both two-step
starting solutions. GAS-NN-1~(30-30) reduces the mean
NLL from 1.2999 to 1.2906, while GAS-NN-1~(200-30)
reduces it from 1.2868 to 1.2858, achieving the lowest
mean NLL among the six FTSE specifications. The larger improvement
from the 30-epoch initializer suggests that joint
estimation provides greater scope for refinement when
the initial neural score has received less training.
By contrast, the 200-epoch initializer already achieves
a relatively low NLL, leaving a smaller improvement
from the subsequent 30 joint updates.

Overall, the results suggest that both additional
score-matching training and joint likelihood estimation
can improve the in-sample fit of GAS-NN models.
The best-performing specification,
GAS-NN-1~(200-30), achieves a mean NLL below
both parametric benchmarks. These findings are
descriptive in-sample comparisons and do not establish
out-of-sample forecasting superiority.

\subsection{Identified negative scores and innovation densities}
\label{identified-innovation-densities-and-negative-scores}

\begin{figure}[!htbp]
\centering
\includegraphics[width=.95\linewidth]{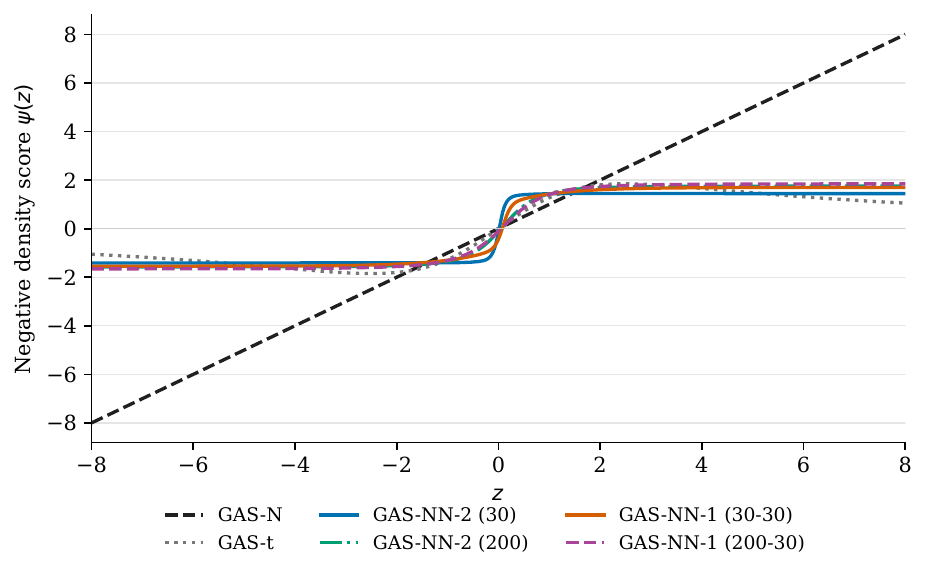}
\caption{FTSE identified negative innovation-density scores}
\label{fig:ftse-negative-scores}
\begin{minipage}{.95\linewidth}
\footnotesize
\textit{Notes:}
The figure displays the negative innovation-density
scores $\phi(z)$ defined in Equation~\eqref{eq:phi-definition},
with the identified neural scores specified in
Equation~\eqref{eq:identified-neural-score}.
The corresponding log-variance updating score
is given in Equation~\eqref{eq:gas-score-phi}.
\end{minipage}
\end{figure}

Figure~\ref{fig:ftse-negative-scores} presents the
identified negative innovation-density scores for
the FTSE market. The negative density score $\phi$
is defined in Equation~\eqref{eq:phi-definition},
while its neural representation $\mathrm{NN}$ is
obtained by relocating and rescaling the raw
neural score $g$, as specified in
Equation~\eqref{eq:identified-neural-score}.
The Gaussian and Student-$t$ scores are given
in Equations~\eqref{eq:gaussian-gas-scores}
and~\eqref{eq:student-t-gas-scores}, respectively.

The Gaussian negative score increases linearly
with the standardized innovation, whereas the
Student-$t$ negative score approaches zero in
the tails. In contrast, the neural negative scores
exhibit nonlinear patterns, with relatively steep
transitions around the centre and gradual
flattening towards the tails.
GAS-NN-2~(30) exhibits a particularly steep
central transition, while GAS-NN-2~(200),
GAS-NN-1~(30-30), and GAS-NN-1~(200-30)
show more gradual transitions. The neural
scores also exhibit differences between their
left and right tail magnitudes, reflecting the
flexibility to accommodate asymmetric
innovation distributions.

\FloatBarrier

The flattening of the neural negative density
scores does not imply a bounded volatility
update. As shown in Equation~\eqref{eq:gas-nn-model},
the log-variance updating score depends on
the product of the standardized innovation
and its identified negative density score.
Since the neural negative scores approach
nonzero constants in the outer tails,
the resulting log-variance score grows
approximately linearly with the shock
magnitude. This growth is slower than the
quadratic response under Gaussian innovations,
but differs from the bounded response under
Student-$t$ innovations.

\begin{figure}[!htbp]
\centering
\includegraphics[width=.95\linewidth]{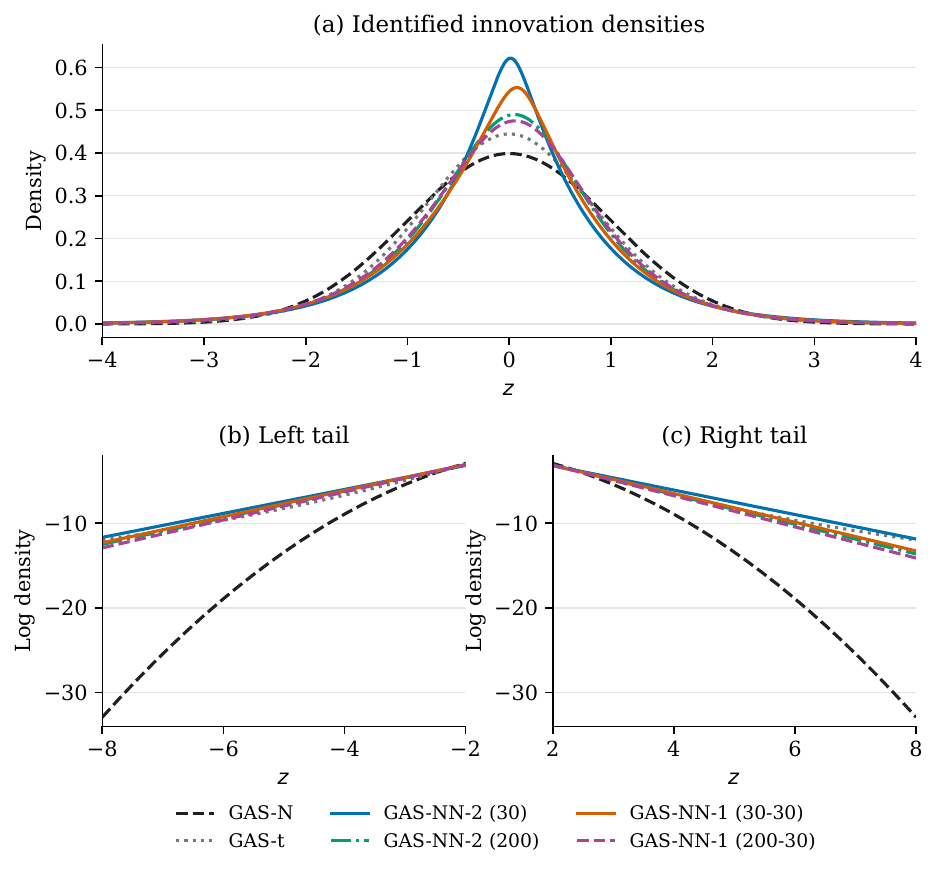}
\caption{FTSE identified innovation densities}
\label{fig:ftse-densities}
\begin{minipage}{.95\linewidth}
\footnotesize
\textit{Notes:}
Panel (a) shows the identified innovation densities.
Panels (b) and (c) show their natural log densities
on a common vertical scale in the left and right
tails. All densities satisfy the identification
conditions in
Equation~\eqref{eq:innovation-identification-conditions}.
Colours and line styles match
Figure~\ref{fig:ftse-negative-scores}.
\end{minipage}
\end{figure}

The identified negative density scores
also determine the shapes of their corresponding
innovation densities. As established in
Section~\ref{estimation-identification},
the raw neural score $g$ determines the
reconstructed density $q$ through
Equation~\eqref{eq:raw-neural-density},
which is subsequently relocated and rescaled
according to Equation~\eqref{eq:identified-neural-density}.
The resulting identified densities $p$ satisfy the
normalization, zero-mean, and unit-variance
conditions in
Equation~\eqref{eq:innovation-identification-conditions}.
Identified negative scores are therefore directly connected
to the identified density shapes displayed
in Figure~\ref{fig:ftse-densities}.

Figure~\ref{fig:ftse-densities} displays the
identified innovation densities and their left
and right log-density tails. GAS-NN-2~(30)
has a particularly concentrated central peak,
consistent with the steep transition in its
negative density score. Longer score matching
under GAS-NN-2~(200) and joint refinement
under GAS-NN-1~(30-30) reduce this central
concentration. GAS-NN-2~(200) and
GAS-NN-1~(200-30) produce similar density
shapes across much of the plotted region,
consistent with their relatively small
difference in in-sample NLL reported in
Table~\ref{tab:ftse}.

\FloatBarrier

The log-tail panels reveal differences that are
less visible on the ordinary density scale.
Across the displayed extreme-return range,
the neural densities allocate substantially
more probability mass than the Gaussian
density. GAS-NN-2~(30) exhibits the slowest
decay among the neural specifications in
the displayed outer tails, while the
longer-trained and jointly refined
specifications show differences in their
left and right decay rates. These patterns
are consistent with the corresponding
negative density scores in
Figure~\ref{fig:ftse-negative-scores}.

As described in
Section~\ref{estimation-identification},
the raw neural score is held constant
outside the reconstruction interval,
subject to the boundary conditions in
Equation~\eqref{eq:neural-boundary-penalty}.
Consequently, the reconstructed and
identified neural densities have
exponential far tails by construction.
The estimated densities therefore allow
flexible and potentially asymmetric
tail behaviour over the plotted range,
without implying power-law decay.

\subsection{Residual quantile diagnostics}
\label{residual-quantile-diagnostics}

\begin{figure}[!htbp]\centering
\includegraphics[width=.86\linewidth]{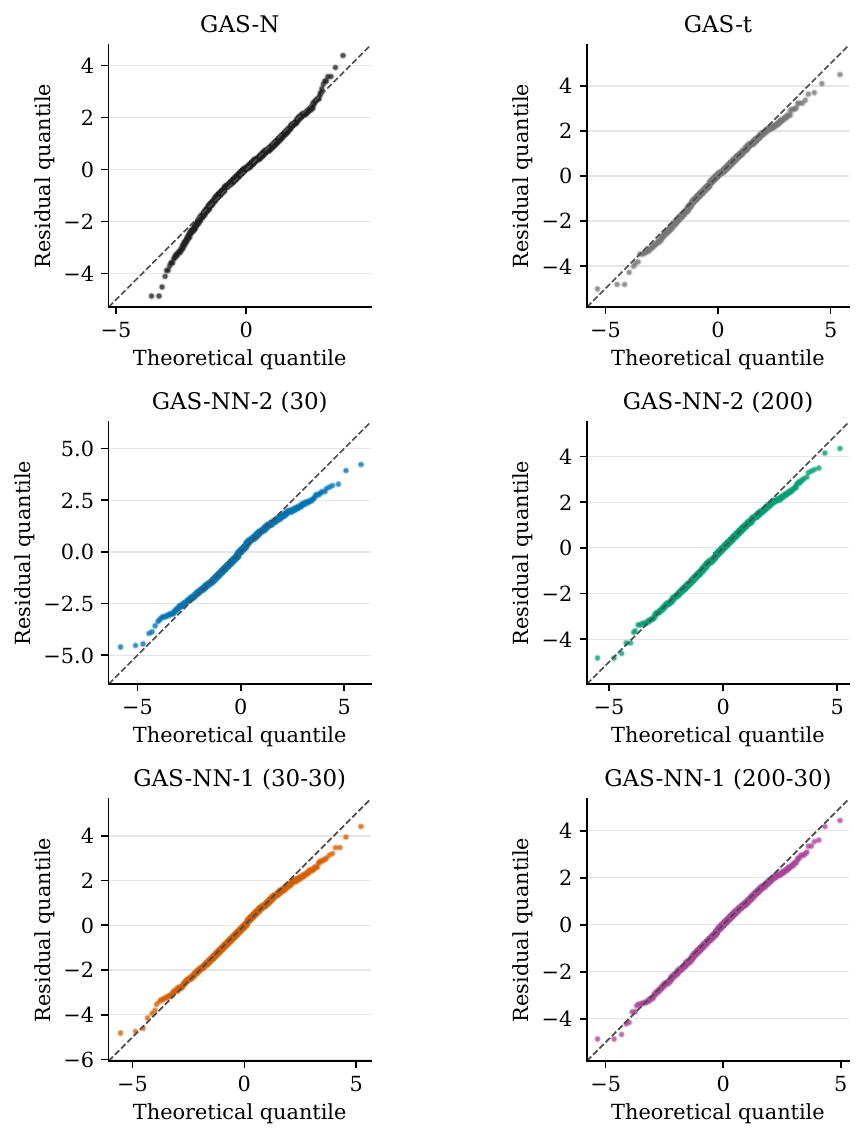}
\caption{FTSE model-specific residual Q--Q plots}\label{fig:3}
\begin{minipage}{.95\linewidth}\footnotesize\textit{Notes:} Horizontal axes show theoretical quantiles from each model's fitted innovation density, while vertical axes show fitted standardized residual quantiles. Dashed lines indicate equality. Each panel uses equal horizontal and vertical scales, although ranges differ across panels.\end{minipage}
\end{figure}

Figure~\ref{fig:3} compares the fitted standardized residual quantiles with the theoretical quantiles of each model's own innovation distribution. The reference distribution is Gaussian for GAS-N, the fitted standardized Student-$t$ distribution for GAS-t, and the identified neural innovation densities for GAS-NN-2 and GAS-NN-1.

\FloatBarrier

The GAS-N panel shows departures in both tails, with extreme residuals extending beyond the corresponding Gaussian quantiles. GAS-t improves tail alignment, although deviations remain among the largest positive quantiles. GAS-NN-2~(30) exhibits pronounced upper-tail departures below the diagonal, indicating that its theoretical positive quantiles are more extreme than the corresponding observed residual quantiles. Increasing the score-matching training budget from 30 to 200 epochs improves the quantile alignment, particularly in the upper tail, as shown by GAS-NN-2~(200). Both GAS-NN-1 specifications further improve the alignment following joint estimation. In particular, GAS-NN-1~(200-30) tracks the reference line more closely than GAS-NN-2~(200) at the extreme positive quantiles, suggesting that the additional joint updates further improve the fitted innovation distribution. Overall, the neural specifications show progressively better quantile alignment with additional training, although discrepancies remain at the extremes.

\section{Out-of-sample analysis}
\label{out-of-sample-analysis}

\subsection{Rolling forecast design}
\label{rolling-forecast-design}

We evaluate one-day-ahead forecasts using a fixed-length rolling
window for each market. The window length equals its initial
pre-2020 in-sample size, ranging from 3,921 observations for
N225 to 4,102 for STOXX50E.
At each forecast origin, the estimation window contains only
observations available before the target date. Its sample mean
and standard deviation are recomputed, and all six models are
re-estimated using the common configuration described in
Section~\ref{sec:architecture_selection}. The fitted models are
filtered through the estimation window to obtain the next
day's conditional variance and predictive distribution. The
window then advances by one trading day, dropping the oldest
observation and adding the newly observed return.

Neural networks and optimizers are initialized afresh at each
forecast origin; estimates are not carried forward from the
preceding window. Within each window, GAS-NN-2~(30) and
GAS-NN-2~(200) use the neural estimates retained after 30 and
200 epochs of the same score-matching trajectory. Each
two-step estimate initializes its corresponding one-step fit,
yielding GAS-NN-1~(30-30) and GAS-NN-1~(200-30) after 30
additional joint updates. The network architecture and training
budgets remain fixed throughout the forecasting period, while
all model parameters are estimated separately for each market
and window.

The full out-of-sample period extends from the first available
trading day of January 2020 to 29 May 2026. We also evaluate
the COVID subsample covering January--June 2020. Differences
in trading calendars produce 1,562--1,641 full-period forecasts
and 119--126 COVID-period forecasts per market. Within each
market and evaluation period, all models are compared on the
same target dates.

\subsection{Forecast evaluation criteria}
\label{forecast-densities-and-loss-functions}

We evaluate forecasting performance using predictive negative
log-likelihood (PNLL), the quantile score (QS), and the
asymmetric-Laplace joint loss (AL). These criteria assess the
predictive density, VaR forecasts, and joint VaR and ES
forecasts, respectively. Let $T_{\mathrm{in}}$ denote the
rolling-window length and $T_{\mathrm{out}}$ the number of
forecasts in an evaluation period. For each period, we index
the evaluated forecasts by
$t=T_{\mathrm{in}}+1,\ldots,T_{\mathrm{in}}+T_{\mathrm{out}}$.

First, we assess predictive density forecasts using PNLL
\citep{gneiting2007}. Following the conditional likelihood
in Equation~\eqref{eq:neural-conditional-nll}, its sample
average is
\begin{equation}
\label{eq:13}
\mathrm{PNLL}
  :=\frac{1}{T_{\mathrm{out}}}
    \sum_{t=T_{\mathrm{in}}+1}^{T_{\mathrm{in}}+T_{\mathrm{out}}}
    \left[
      \frac{\widehat f_t}{2}-\log p(z_t)
    \right],
\end{equation}
where $\widehat f_t$ is the one-day-ahead conditional
log-variance forecast and
$z_t=r_t\exp(-\widehat f_t/2)$ is the standardized
forecast residual. Here, $r_t$ is demeaned using the
corresponding rolling-window mean, and $p$ is the fitted
innovation density for that window. Its dependence on
the rolling parameter estimates is suppressed for
notational simplicity. PNLL therefore evaluates each
model's own predictive density.

Second, we evaluate one-day-ahead VaR and ES forecasts,
denoted by $\widehat{\mathrm{VaR}}_t$ and
$\widehat{\mathrm{ES}}_t$, at the $\delta =2.5\%$ probability level.
VaR is the conditional 2.5\% quantile, and ES is the
conditional expected return below that quantile.

VaR forecasts are evaluated using the consistent quantile
score \citep{gneiting2011}, averaged over the evaluation
period:
\begin{equation}
\label{eq:15}
\mathrm{QS}
  :=\frac{1}{T_{\mathrm{out}}}
    \sum_{t=T_{\mathrm{in}}+1}^{T_{\mathrm{in}}+T_{\mathrm{out}}}
    \left(
      \delta-I\{r_t\leq\widehat{\mathrm{VaR}}_t\}
    \right)
    \left(
      r_t-\widehat{\mathrm{VaR}}_t
    \right),
\end{equation}
where $I\{\cdot\}$ equals one when the condition holds
and zero otherwise.

Third, although ES is not individually elicitable, it is
jointly elicitable with VaR \citep{gneiting2011,fissler2016}.
Following \citet{taylor2019}, we evaluate the paired
forecasts using the AL joint loss:
\begin{equation}
\label{eq:16}
\mathrm{AL}
  :=\frac{1}{T_{\mathrm{out}}}
    \sum_{t=T_{\mathrm{in}}+1}^{T_{\mathrm{in}}+T_{\mathrm{out}}}
    \left[
      -\log\left(
        \frac{\delta-1}{\widehat{\mathrm{ES}}_t}
      \right)
      -\frac{
        (r_t-\widehat{\mathrm{VaR}}_t)
        (\delta-I\{r_t\leq\widehat{\mathrm{VaR}}_t\})
      }{
        \delta\,\widehat{\mathrm{ES}}_t
      }
    \right].
\end{equation}
All three criteria are averaged over the corresponding
evaluation period, and lower values indicate better
forecast performance.

We also use the model confidence set (MCS) procedure of
\citet{hansen2011} to assess the uncertainty surrounding
the numerical loss rankings. A separate MCS is constructed
for each market, evaluation period, and loss function using
the daily losses underlying PNLL, QS, and AL. The confidence level is 75\%, and models
within each comparison share the same bootstrap draws.

% In Tables~\ref{tab:4}--\ref{tab:6}, blue entries identify
% the lowest mean loss within each market and panel, and grey
% shading denotes membership of the corresponding 75\% MCS. MCS membership indicates that a model
% has not been eliminated by the procedure; it does not
% establish that the model is uniquely best. Counts across
% markets are descriptive, since market outcomes are
% dependent and the COVID subsample is contained within the
% full evaluation period.

\subsection{Predictive density performance}
\label{predictive-density-performance}

Table~\ref{tab:4} reports PNLL. Over the full period,
GAS-NN-1~(200-30) achieves the lowest PNLL for the S\&P 500,
AORD, and N225; GAS-NN-2~(200) ranks first for GDAXI,
DJI, and STOXX50E; and GAS-NN-1~(30-30) ranks
first for FTSE. GAS-t performs best for HSI. Neural
specifications therefore attain the lowest PNLL in seven
of the eight markets, although joint estimation does not
uniformly improve on the corresponding two-step fit.

The full-period MCS retains GAS-NN-2~(200) and
GAS-NN-1~(200-30) in six markets each, GAS-NN-1~(30-30)
in five, and GAS-t in three. GAS-N and GAS-NN-2~(30)
are excluded in all eight markets. These results favour
the longer score-matching budget for full-period density
forecasting, while the benefit of subsequent joint
estimation varies across markets.

During the COVID period, GAS-NN-1~(30-30) achieves the
lowest PNLL in all eight markets, ranging from 1.9381
for HSI to 2.2183 for DJI. GAS-NN-2~(200),
GAS-NN-1~(30-30), and GAS-NN-1~(200-30) are retained
in every market's MCS. The consistent numerical ranking
of GAS-NN-1~(30-30) therefore indicates strong crisis-period
density performance, while the MCS results leave several
neural alternatives statistically competitive.

\subsection{Quantile forecast performance}
\label{quantile-forecast-performance}

Table~\ref{tab:5} reports QS at the 2.5\% level.
Over the full period, GAS-NN-1~(30-30) ranks first for
the S\&P 500, FTSE, GDAXI, DJI, and STOXX50E.
GAS-NN-2~(30) achieves the lowest QS for AORD, GAS-t
for HSI, and GAS-N for N225. During the COVID
period, GAS-NN-1~(30-30) remains first for the S\&P 500,
FTSE, DJI, and STOXX50E; GAS-N ranks first
for GDAXI and AORD; and GAS-NN-2~(30) ranks first for
HSI and N225. The quantile rankings therefore
differ from those for the complete predictive density.

The MCS results indicate limited separation between
models under QS. GAS-t is retained in all eight
full-period comparisons, while GAS-N, GAS-NN-2~(30),
GAS-NN-2~(200), and GAS-NN-1~(30-30) are each retained
in seven. GAS-NN-1~(200-30) is retained in five.
During COVID, GAS-NN-2~(30) is retained in all eight
markets, and multiple competing specifications remain
in most confidence sets. Thus, GAS-NN-1~(30-30) achieves
the largest number of minimum-loss entries in both
periods, but these numerical gains do not establish
uniform statistical superiority.

\subsection{Joint VaR and ES forecast performance}
\label{joint-var-and-es-forecast-performance}

Table~\ref{tab:6} reports AL at the 2.5\% level.
Over the full period, GAS-NN-2~(30) achieves the lowest
AL in six markets: the S\&P 500, FTSE, GDAXI, AORD,
DJI, and STOXX50E. GAS-t ranks first for
HSI, while GAS-NN-1~(30-30) ranks first for N225.
GAS-NN-2~(30) and GAS-NN-1~(30-30) are each retained
in seven full-period MCS comparisons.

During the COVID period, GAS-NN-2~(30) achieves the
lowest AL in all eight markets, with mean losses
ranging from 2.7203 for N225 to 3.4616 for GDAXI.
It is retained in all eight MCS comparisons. GAS-N
is also retained in seven markets, and both GAS-NN-1
specifications are retained in six. The uniform
minimum-loss ranking of GAS-NN-2~(30) therefore
coexists with substantial uncertainty about the
differences between competing forecasts.

The contrast between PNLL and AL shows that the gains
from neural estimation depend on the forecast target.
GAS-NN-2~(30) is excluded from every full-period PNLL
confidence set but performs strongly under AL.
GAS-NN-1~(30-30), meanwhile, leads the COVID density
comparison. Joint likelihood estimation targets the
complete conditional distribution, so improvements
in overall density fit need not produce the lowest
realized quantile or joint tail-risk loss.

\begin{table}[htbp]
\centering
\caption{Mean predictive negative log-likelihood (PNLL)}
\label{tab:4}
\small
\setlength{\tabcolsep}{3pt}
\renewcommand{\arraystretch}{1.1}
\begin{tabular}{@{}lZZZZ:ZZ@{}}
\hline
Full period
& \multicolumn{4}{c:}{Proposed Models}
& \multicolumn{2}{c}{Benchmarks} \\
01/01/2020--29/05/2026
& \shortstack{GAS-NN-2\\(30)}
& \shortstack{GAS-NN-2\\(200)}
& \shortstack{GAS-NN-1\\(30-30)}
& \shortstack{GAS-NN-1\\(200-30)}
& GAS-N & GAS-t \\
\hline
S\&P 500
& 1.4236
& \cellcolor{gray!25}1.4128
& 1.4153
& \cellcolor{gray!25}\textcolor{blue}{1.4126}
& 1.4373
& \cellcolor{gray!25}1.4145 \\
FTSE
& 1.2131
& \cellcolor{gray!25}1.2095
& \cellcolor{gray!25}\textcolor{blue}{1.2078}
& \cellcolor{gray!25}1.2101
& 1.2635
& 1.2159 \\
GDAXI
& 1.4655
& \cellcolor{gray!25}\textcolor{blue}{1.4582}
& \cellcolor{gray!25}1.4604
& \cellcolor{gray!25}1.4585
& 1.4984
& 1.4626 \\
AORD
& 1.2508
& 1.2344
& \cellcolor{gray!25}1.2336
& \cellcolor{gray!25}\textcolor{blue}{1.2329}
& 1.2541
& 1.2402 \\
DJI
& 1.3504
& \cellcolor{gray!25}\textcolor{blue}{1.3411}
& \cellcolor{gray!25}1.3433
& \cellcolor{gray!25}1.3415
& 1.3639
& 1.3437 \\
HSI
& 1.8043
& 1.7947
& 1.7961
& 1.7936
& 1.8401
& \cellcolor{gray!25}\textcolor{blue}{1.7883} \\
N225
& 1.6689
& \cellcolor{gray!25}1.6612
& \cellcolor{gray!25}1.6620
& \cellcolor{gray!25}\textcolor{blue}{1.6611}
& 1.6764
& \cellcolor{gray!25}1.6628 \\
STOXX50E
& 1.4778
& \cellcolor{gray!25}\textcolor{blue}{1.4704}
& 1.4726
& 1.4710
& 1.5117
& 1.4739 \\
\hline
COVID period
& \multicolumn{4}{c:}{Proposed Models}
& \multicolumn{2}{c}{Benchmarks} \\
01/01/2020--30/06/2020
& \shortstack{GAS-NN-2\\(30)}
& \shortstack{GAS-NN-2\\(200)}
& \shortstack{GAS-NN-1\\(30-30)}
& \shortstack{GAS-NN-1\\(200-30)}
& GAS-N & GAS-t \\
\hline
S\&P 500
& 2.1458
& \cellcolor{gray!25}2.1324
& \cellcolor{gray!25}\textcolor{blue}{2.1257}
& \cellcolor{gray!25}2.1304
& \cellcolor{gray!25}2.2175
& \cellcolor{gray!25}2.1386 \\
FTSE
& 2.0554
& \cellcolor{gray!25}2.0435
& \cellcolor{gray!25}\textcolor{blue}{2.0387}
& \cellcolor{gray!25}2.0437
& 2.1032
& 2.0595 \\
GDAXI
& \cellcolor{gray!25}2.2045
& \cellcolor{gray!25}2.2101
& \cellcolor{gray!25}\textcolor{blue}{2.2004}
& \cellcolor{gray!25}2.2109
& \cellcolor{gray!25}2.2845
& 2.2284 \\
AORD
& 2.0289
& \cellcolor{gray!25}2.0183
& \cellcolor{gray!25}\textcolor{blue}{1.9971}
& \cellcolor{gray!25}2.0154
& \cellcolor{gray!25}2.0367
& 2.0376 \\
DJI
& 2.2326
& \cellcolor{gray!25}2.2278
& \cellcolor{gray!25}\textcolor{blue}{2.2183}
& \cellcolor{gray!25}2.2269
& \cellcolor{gray!25}2.3194
& \cellcolor{gray!25}2.2353 \\
HSI
& \cellcolor{gray!25}1.9463
& \cellcolor{gray!25}1.9427
& \cellcolor{gray!25}\textcolor{blue}{1.9381}
& \cellcolor{gray!25}1.9424
& 1.9847
& 1.9540 \\
N225
& \cellcolor{gray!25}2.0880
& \cellcolor{gray!25}2.0826
& \cellcolor{gray!25}\textcolor{blue}{2.0805}
& \cellcolor{gray!25}2.0828
& \cellcolor{gray!25}2.0827
& \cellcolor{gray!25}2.0936 \\
STOXX50E
& \cellcolor{gray!25}2.1474
& \cellcolor{gray!25}2.1476
& \cellcolor{gray!25}\textcolor{blue}{2.1419}
& \cellcolor{gray!25}2.1499
& 2.2233
& 2.1632 \\
\hline
\end{tabular}
\captionsetup{
    font=footnotesize,
    justification=raggedright,
    singlelinecheck=false
}
\caption*{\textit{Notes:} Lower values indicate better forecasting performance.
Blue values identify the lowest loss for each market and period, based on
unrounded values; all reported losses are rounded to four decimal places.
Grey shading indicates retention in the 75\% MCS (range statistic;
stationary bootstrap with 10,000 replications, expected block length 10,
and seed 1234). Periods include available trading observations within the
dates shown. Sample sizes are reported in Table~\ref{tab:samples}.
Each model is evaluated using its own complete conditional density.
GAS-NN-2 uses two-step estimation with 30 or 200 score-matching epochs;
GAS-NN-1 uses the corresponding two-step initializer followed by 30 joint
updates. GAS-N and GAS-t are the Gaussian and standardized Student-$t$
benchmarks.}
\end{table}

\begin{table}[htbp]
\centering
\caption{Mean quantile score (QS) at the 2.5\% level}
\label{tab:5}
\small
\setlength{\tabcolsep}{3pt}
\renewcommand{\arraystretch}{1.1}
\begin{tabular}{@{}lZZZZ:ZZ@{}}
\hline
Full period
& \multicolumn{4}{c:}{Proposed Models}
& \multicolumn{2}{c}{Benchmarks} \\
01/01/2020--29/05/2026
& \shortstack{GAS-NN-2\\(30)}
& \shortstack{GAS-NN-2\\(200)}
& \shortstack{GAS-NN-1\\(30-30)}
& \shortstack{GAS-NN-1\\(200-30)}
& GAS-N & GAS-t \\
\hline
S\&P 500 & \cellcolor{gray!25}0.0832 & \cellcolor{gray!25}0.0833 & \cellcolor{gray!25}\textcolor{blue}{0.0825} & \cellcolor{gray!25}0.0834 & \cellcolor{gray!25}0.0844 & \cellcolor{gray!25}0.0838 \\
FTSE & \cellcolor{gray!25}0.0731 & \cellcolor{gray!25}0.0735 & \cellcolor{gray!25}\textcolor{blue}{0.0731} & 0.0737 & \cellcolor{gray!25}0.0733 & \cellcolor{gray!25}0.0742 \\
GDAXI & \cellcolor{gray!25}0.0902 & \cellcolor{gray!25}0.0905 & \cellcolor{gray!25}\textcolor{blue}{0.0902} & \cellcolor{gray!25}0.0905 & \cellcolor{gray!25}0.0909 & \cellcolor{gray!25}0.0910 \\
AORD & \cellcolor{gray!25}\textcolor{blue}{0.0705} & \cellcolor{gray!25}0.0716 & \cellcolor{gray!25}0.0707 & \cellcolor{gray!25}0.0717 & \cellcolor{gray!25}0.0710 & \cellcolor{gray!25}0.0729 \\
DJI & \cellcolor{gray!25}0.0775 & \cellcolor{gray!25}0.0776 & \cellcolor{gray!25}\textcolor{blue}{0.0768} & \cellcolor{gray!25}0.0777 & \cellcolor{gray!25}0.0779 & \cellcolor{gray!25}0.0783 \\
HSI & 0.1039 & 0.1028 & 0.1028 & 0.1024 & 0.1058 & \cellcolor{gray!25}\textcolor{blue}{0.1011} \\
N225 & \cellcolor{gray!25}0.0973 & \cellcolor{gray!25}0.0972 & \cellcolor{gray!25}0.0972 & \cellcolor{gray!25}0.0973 & \cellcolor{gray!25}\textcolor{blue}{0.0968} & \cellcolor{gray!25}0.0973 \\
STOXX50E & \cellcolor{gray!25}0.0900 & \cellcolor{gray!25}0.0904 & \cellcolor{gray!25}\textcolor{blue}{0.0899} & 0.0906 & \cellcolor{gray!25}0.0907 & \cellcolor{gray!25}0.0908 \\
\hline
COVID period
& \multicolumn{4}{c:}{Proposed Models}
& \multicolumn{2}{c}{Benchmarks} \\
01/01/2020--30/06/2020
& \shortstack{GAS-NN-2\\(30)}
& \shortstack{GAS-NN-2\\(200)}
& \shortstack{GAS-NN-1\\(30-30)}
& \shortstack{GAS-NN-1\\(200-30)}
& GAS-N & GAS-t \\
\hline
S\&P 500 & \cellcolor{gray!25}0.2127 & \cellcolor{gray!25}0.2235 & \cellcolor{gray!25}\textcolor{blue}{0.2065} & \cellcolor{gray!25}0.2215 & \cellcolor{gray!25}0.2405 & \cellcolor{gray!25}0.2293 \\
FTSE & \cellcolor{gray!25}0.2030 & \cellcolor{gray!25}0.2099 & \cellcolor{gray!25}\textcolor{blue}{0.2028} & \cellcolor{gray!25}0.2102 & \cellcolor{gray!25}0.2083 & \cellcolor{gray!25}0.2149 \\
GDAXI & \cellcolor{gray!25}0.2397 & \cellcolor{gray!25}0.2446 & \cellcolor{gray!25}0.2386 & \cellcolor{gray!25}0.2450 & \cellcolor{gray!25}\textcolor{blue}{0.2373} & \cellcolor{gray!25}0.2492 \\
AORD & \cellcolor{gray!25}0.1969 & \cellcolor{gray!25}0.2075 & \cellcolor{gray!25}0.1946 & \cellcolor{gray!25}0.2078 & \cellcolor{gray!25}\textcolor{blue}{0.1869} & \cellcolor{gray!25}0.2174 \\
DJI & \cellcolor{gray!25}0.2309 & \cellcolor{gray!25}0.2418 & \cellcolor{gray!25}\textcolor{blue}{0.2249} & \cellcolor{gray!25}0.2399 & \cellcolor{gray!25}0.2573 & \cellcolor{gray!25}0.2467 \\
HSI & \cellcolor{gray!25}\textcolor{blue}{0.1375} & 0.1475 & 0.1404 & 0.1482 & 0.1572 & 0.1548 \\
N225 & \cellcolor{gray!25}\textcolor{blue}{0.1371} & \cellcolor{gray!25}0.1462 & \cellcolor{gray!25}0.1401 & 0.1474 & \cellcolor{gray!25}0.1449 & 0.1526 \\
STOXX50E & \cellcolor{gray!25}0.2367 & \cellcolor{gray!25}0.2451 & \cellcolor{gray!25}\textcolor{blue}{0.2360} & \cellcolor{gray!25}0.2453 & \cellcolor{gray!25}0.2366 & \cellcolor{gray!25}0.2513 \\
\hline
\end{tabular}
\captionsetup{
    font=footnotesize,
    justification=raggedright,
    singlelinecheck=false
}
\caption*{\textit{Notes:} Lower values indicate better forecasting performance.
Blue values identify the lowest loss for each market and period, based on
unrounded values; all reported losses are rounded to four decimal places.
Grey shading indicates retention in the 75\% MCS (range statistic;
stationary bootstrap with 10,000 replications, expected block length 10,
and seed 1234). Periods include available trading observations within the
dates shown. Sample sizes are reported in Table~\ref{tab:samples}.
GAS-NN-2 uses two-step estimation with 30 or 200 score-matching epochs;
GAS-NN-1 uses the corresponding two-step initializer followed by 30 joint
updates. GAS-N and GAS-t are the Gaussian and standardized Student-$t$
benchmarks.}
\end{table}

\begin{table}[htbp]
\centering
\caption{Mean AL joint loss at the 2.5\% level}
\label{tab:6}
\small
\setlength{\tabcolsep}{3pt}
\renewcommand{\arraystretch}{1.1}
\begin{tabular}{@{}lZZZZ:ZZ@{}}
\hline
Full period
& \multicolumn{4}{c:}{Proposed Models}
& \multicolumn{2}{c}{Benchmarks} \\
01/01/2020--29/05/2026
& \shortstack{GAS-NN-2\\(30)}
& \shortstack{GAS-NN-2\\(200)}
& \shortstack{GAS-NN-1\\(30-30)}
& \shortstack{GAS-NN-1\\(200-30)}
& GAS-N & GAS-t \\
\hline
S\&P 500 & \cellcolor{gray!25}\textcolor{blue}{2.1782} & \cellcolor{gray!25}2.1950 & \cellcolor{gray!25}2.1845 & \cellcolor{gray!25}2.1895 & \cellcolor{gray!25}2.2374 & \cellcolor{gray!25}2.2132 \\
FTSE & \cellcolor{gray!25}\textcolor{blue}{2.0063} & \cellcolor{gray!25}2.0211 & \cellcolor{gray!25}2.0114 & 2.0288 & 2.0598 & 2.0476 \\
GDAXI & \cellcolor{gray!25}\textcolor{blue}{2.2493} & \cellcolor{gray!25}2.2628 & \cellcolor{gray!25}2.2559 & \cellcolor{gray!25}2.2627 & \cellcolor{gray!25}2.2973 & \cellcolor{gray!25}2.2835 \\
AORD & \cellcolor{gray!25}\textcolor{blue}{1.9727} & 1.9931 & 1.9803 & 1.9980 & 2.0456 & 2.0392 \\
DJI & \cellcolor{gray!25}\textcolor{blue}{2.0595} & \cellcolor{gray!25}2.0668 & \cellcolor{gray!25}2.0601 & \cellcolor{gray!25}2.0649 & \cellcolor{gray!25}2.0939 & \cellcolor{gray!25}2.0851 \\
HSI & 2.4148 & 2.4021 & \cellcolor{gray!25}2.4001 & \cellcolor{gray!25}2.3979 & \cellcolor{gray!25}2.4251 & \cellcolor{gray!25}\textcolor{blue}{2.3859} \\
N225 & \cellcolor{gray!25}2.3238 & \cellcolor{gray!25}2.3240 & \cellcolor{gray!25}\textcolor{blue}{2.3207} & \cellcolor{gray!25}2.3255 & \cellcolor{gray!25}2.3314 & \cellcolor{gray!25}2.3325 \\
STOXX50E & \cellcolor{gray!25}\textcolor{blue}{2.2472} & \cellcolor{gray!25}2.2630 & \cellcolor{gray!25}2.2537 & 2.2665 & \cellcolor{gray!25}2.2919 & \cellcolor{gray!25}2.2794 \\
\hline
COVID period
& \multicolumn{4}{c:}{Proposed Models}
& \multicolumn{2}{c}{Benchmarks} \\
01/01/2020--30/06/2020
& \shortstack{GAS-NN-2\\(30)}
& \shortstack{GAS-NN-2\\(200)}
& \shortstack{GAS-NN-1\\(30-30)}
& \shortstack{GAS-NN-1\\(200-30)}
& GAS-N & GAS-t \\
\hline
S\&P 500 & \cellcolor{gray!25}\textcolor{blue}{3.2559} & 3.4178 & \cellcolor{gray!25}3.2591 & \cellcolor{gray!25}3.3734 & \cellcolor{gray!25}3.7014 & \cellcolor{gray!25}3.5290 \\
FTSE & \cellcolor{gray!25}\textcolor{blue}{3.1319} & \cellcolor{gray!25}3.2767 & \cellcolor{gray!25}3.1558 & \cellcolor{gray!25}3.2973 & \cellcolor{gray!25}3.4243 & \cellcolor{gray!25}3.4197 \\
GDAXI & \cellcolor{gray!25}\textcolor{blue}{3.4616} & \cellcolor{gray!25}3.5962 & \cellcolor{gray!25}3.4764 & \cellcolor{gray!25}3.6054 & \cellcolor{gray!25}3.6317 & \cellcolor{gray!25}3.7353 \\
AORD & \cellcolor{gray!25}\textcolor{blue}{3.1077} & \cellcolor{gray!25}3.2845 & \cellcolor{gray!25}3.1213 & \cellcolor{gray!25}3.3032 & \cellcolor{gray!25}3.2682 & 3.4919 \\
DJI & \cellcolor{gray!25}\textcolor{blue}{3.3549} & \cellcolor{gray!25}3.5104 & \cellcolor{gray!25}3.3699 & \cellcolor{gray!25}3.4716 & \cellcolor{gray!25}3.7837 & \cellcolor{gray!25}3.6264 \\
HSI & \cellcolor{gray!25}\textcolor{blue}{2.7611} & 2.8890 & 2.8078 & 2.9031 & 3.1000 & 3.0061 \\
N225 & \cellcolor{gray!25}\textcolor{blue}{2.7203} & 2.8413 & 2.7683 & 2.8572 & \cellcolor{gray!25}2.9001 & 2.9472 \\
STOXX50E & \cellcolor{gray!25}\textcolor{blue}{3.3937} & \cellcolor{gray!25}3.5641 & \cellcolor{gray!25}3.4189 & \cellcolor{gray!25}3.5831 & \cellcolor{gray!25}3.6153 & \cellcolor{gray!25}3.6997 \\
\hline
\end{tabular}
\captionsetup{
    font=footnotesize,
    justification=raggedright,
    singlelinecheck=false
}
\caption*{\textit{Notes:} Lower values indicate better forecasting performance.
Blue values identify the lowest loss for each market and period, based on
unrounded values; all reported losses are rounded to four decimal places.
Grey shading indicates retention in the 75\% MCS (range statistic;
stationary bootstrap with 10,000 replications, expected block length 10,
and seed 1234). Periods include available trading observations within the
dates shown. Sample sizes are reported in Table~\ref{tab:samples}.
AL is defined in equation~\eqref{eq:16}.
GAS-NN-2 uses two-step estimation with 30 or 200 score-matching epochs;
GAS-NN-1 uses the corresponding two-step initializer followed by 30 joint
updates. GAS-N and GAS-t are the Gaussian and standardized Student-$t$
benchmarks.}
\end{table}

\clearpage
\section{Conclusion}
\label{conclusion}

This paper develops GAS-NN, a volatility model in which a neural network represents the negative innovation-density score within a GAS recursion. Density reconstruction, normalization, and location--scale identification ensure that the learned score corresponds to an innovation density with zero mean and unit variance. The same density determines the volatility update, conditional likelihood, and predictive risk measures. Two-step estimation combines regularized score matching with fixed-density likelihood estimation of the volatility parameters. One-step estimation then permits joint adjustment of the neural density and volatility dynamics through a regularized conditional likelihood.

The application to eight international equity indices shows that a compact architecture selected on the S\&P~500 can be applied across markets with separately estimated parameters. The FTSE results illustrate how longer score matching and joint refinement change the fitted density, score function, and residual quantile alignment. Neural models attain the lowest full-period predictive negative log-likelihood in seven markets. During the first half of 2020, GAS-NN-1~(30-30) attains the lowest predictive negative log-likelihood in all eight markets, whereas GAS-NN-2~(30) attains the lowest joint VaR--ES loss in all eight. The model confidence sets retain several competitors, particularly for tail-risk losses, so these numerical rankings do not establish uniform statistical superiority.

The findings indicate that the amount of first-stage training and the subsequent joint refinement have different implications for density and tail-risk forecasts. Likelihood refinement adjusts the complete predictive distribution, but it does not necessarily minimize realized losses focused on the lower tail. The two-step specification remains useful when its fixed working density produces favourable risk forecasts, even when another model provides a better average density fit.

Several extensions follow from the present design. First-stage training could use preliminary filtered residuals to reduce the influence of time-varying volatility on the score-matching inputs. Alternative tail continuations and objectives focused on risk forecasts could broaden the distributional and forecasting scope of the model. Manifold learning could also be explored to capture lower-dimensional structures in multivariate innovations, extending the neural score framework to multivariate volatility modelling. Further theoretical work could establish stability and estimation properties for the neural specification. The current results provide empirical support for learning innovation-density scores within a structured volatility recursion, with gains that depend on the market, evaluation period, and forecast target.

\clearpage
\bibliographystyle{apacite}
\bibliography{references}

@article{gneiting2011,
  author  = {Gneiting, Tilmann},
  title   = {Making and Evaluating Point Forecasts},
  journal = {Journal of the American Statistical Association},
  year    = {2011},
  volume  = {106},
  number  = {494},
  pages   = {746--762}
}

@misc{who2026,
  author = {{World Health Organization}},
  title = {Archived: {WHO} Timeline -- {COVID}-19},
  year = {2020},
  month = apr,
  url = {https://www.who.int/news/item/27-04-2020-who-timeline---covid-19},
  note = {Accessed 3 July 2026}
}

@article{blasques2016,
  author = {Blasques, Francisco and Ji, Jiangyu and Lucas, Andr{\'e}},
  title = {Semiparametric score driven volatility models},
  journal = {Computational Statistics \& Data Analysis},
  volume = {100}, pages = {58--69}, year = {2016},
  doi = {10.1016/j.csda.2015.04.003}
}

@article{creal2013,
  author = {Creal, Drew and Koopman, Siem Jan and Lucas, Andr{\'e}},
  title = {Generalized Autoregressive Score Models with Applications},
  journal = {Journal of Applied Econometrics},
  volume = {28}, number = {5}, pages = {777--795}, year = {2013},
  doi = {10.1002/jae.1279}
}

@article{hansen2011,
  author = {Hansen, Peter R. and Lunde, Asger and Nason, James M.},
  title = {The Model Confidence Set}, journal = {Econometrica},
  volume = {79}, number = {2}, pages = {453--497}, year = {2011},
  doi = {10.3982/ECTA5771}
}

@article{hyvarinen2005,
  author = {Hyv{\"a}rinen, Aapo},
  title = {Estimation of Non-Normalized Statistical Models by Score Matching},
  journal = {Journal of Machine Learning Research},
  volume = {6}, pages = {695--709}, year = {2005}
}

@article{taylor2019,
  author = {Taylor, James W.},
  title = {Forecasting Value at Risk and Expected Shortfall Using a Semiparametric Approach Based on the Asymmetric Laplace Distribution},
  journal = {Journal of Business \& Economic Statistics},
  volume = {37}, number = {1}, pages = {121--133}, year = {2019},
  doi = {10.1080/07350015.2017.1281815}
}

@article{vanbrummelen2026,
  author = {van Brummelen, Janneke and Gorgi, Paolo and Koopman, Siem Jan},
  title = {Score-driven time-varying parameter models with spline-based densities},
  journal = {Statistics and Computing},
  volume = {36}, pages = {94}, year = {2026},
  doi = {10.1007/s11222-026-10840-w}
}

@article{engle1982,
  author = {Engle, Robert F.},
  title = {Autoregressive Conditional Heteroscedasticity with Estimates of the Variance of {United Kingdom} Inflation},
  journal = {Econometrica},
  year = {1982},
  volume = {50},
  number = {4},
  pages = {987--1007},
  doi = {10.2307/1912773}
}

@article{bollerslev1986,
  author = {Bollerslev, Tim},
  title = {Generalized Autoregressive Conditional Heteroskedasticity},
  journal = {Journal of Econometrics},
  year = {1986},
  volume = {31},
  number = {3},
  pages = {307--327},
  doi = {10.1016/0304-4076(86)90063-1}
}

@book{harvey2013,
  author = {Harvey, Andrew C.},
  title = {Dynamic Models for Volatility and Heavy Tails: With Applications to Financial and Economic Time Series},
  publisher = {Cambridge University Press},
  year = {2013},
  series = {Econometric Society Monographs},
  number = {52},
  isbn = {9781107034723},
  url = {https://www.cambridge.org/core/books/dynamic-models-for-volatility-and-heavy-tails/896F9D5220C4DD2CA675846F888F0BF0}
}

@article{gneiting2007,
  author = {Gneiting, Tilmann and Raftery, Adrian E.},
  title = {Strictly Proper Scoring Rules, Prediction, and Estimation},
  journal = {Journal of the American Statistical Association},
  year = {2007},
  volume = {102},
  number = {477},
  pages = {359--378},
  doi = {10.1198/016214506000001437}
}

@article{fissler2016,
  author = {Fissler, Tobias and Ziegel, Johanna F.},
  title = {Higher Order Elicitability and {Osband}'s Principle},
  journal = {The Annals of Statistics},
  year = {2016},
  volume = {44},
  number = {4},
  pages = {1680--1707},
  doi = {10.1214/16-AOS1439}
}
\clearpage
\appendix
\section{Implementation details}
\label{app:implementation}

The GAS parameter transformations are
$\alpha=\operatorname{softplus}(a_{\mathrm{raw}})$ and
$\beta=0.999\operatorname{sigmoid}(b_{\mathrm{raw}})$, with
$\omega$ represented directly. They enforce $\alpha>0$ and
$0<\beta<0.999$. The initial log variance is
$f_1=\log\{T_{\mathrm{in}}^{-1}\sum_{t=1}^{T_{\mathrm{in}}}r_t^2\}$.
Fixed-density likelihood optimization uses three initial values,
at most 1,000 L-BFGS-B iterations, up to 40 line-search steps,
function tolerance $10^{-10}$, and gradient tolerance $10^{-6}$.
The optimization coordinates $(\omega,a_{\mathrm{raw}},b_{\mathrm{raw}})$
are bounded by $[-5,5]$, $[-12,5]$, and $[-10,12]$, respectively.
For GAS-t, $\nu=2.05+\operatorname{softplus}(v)$ with $v\in[-5,100]$.
These box bounds are not applied during joint Adam updates,
which retain the GAS parameter transformations.

Neural training uses full-batch Adam with learning rate $10^{-3}$
and random seed 1234. Hidden-layer weights use Xavier-uniform
initialization; biases and output-layer weights are initialized
at zero. Score matching uses single precision and joint estimation
uses double precision. The regularized objectives are given in
Equations~\eqref{eq:two-step-procedure} and~\eqref{eq:one-step-procedure},
with $\lambda_B=0.1$ and $\lambda_W=5\times10^{-5}$.
The latter corresponds to coupled neural weight decay $10^{-4}$,
whose contribution is $10^{-4}\|\theta_{\mathrm{NN}}\|_2^2/2$.
This is the same $L_2$ penalty already included in the stated
objectives; the GAS parameters are unpenalized.
Reported likelihoods exclude both regularization penalties.
Fixed-density optimization uses total NLL, which has the same
minimizer as mean NLL for a fixed estimation sample.

Density reconstruction in Section~\ref{estimation-identification}
uses 40,001 equally spaced points on $[-40,40]$, so $B=40$ and
the grid spacing is 0.002. The interior antiderivative and density
moments are approximated by trapezoidal integration, with analytical
contributions from the exponentially decaying outer tails.
Log densities are linearly interpolated between grid points.
Fixed-density filtering evaluates the neural score directly,
whereas joint filtering uses linear interpolation of the score.
The density and score equations specify the continuous model;
these operations provide its numerical approximation.

Risk calculations integrate the log-linear density within each
grid interval and use the analytical outer tails. A discrepancy
in numerical total mass is accepted only within the configured
tolerance, and the risk integrals are divided by their measured
total mass. This quadrature correction does not replace or rescale
the predictive likelihood used in the PNLL comparison.

Numerical safeguards include conditional-variance bounds
$[10^{-8},10^8]$, a tail-rate floor of $10^{-8}$, and a raw-variance
floor of $10^{-12}$. A non-finite fixed-density trial objective
is assigned the value $10^{100}$. Final identified density moments
are checked to tolerance $10^{-5}$, and paired two-step and joint
initialization checks compare the density and NLL evaluations to
tolerance $2\times10^{-7}$. For the final FTSE neural fits, the
recorded mass, mean, and variance checks pass, boundary penalties
are zero, and the tail-rate, raw-variance, and conditional-variance
safeguards have no recorded hits. These checks establish numerical
consistency of the fitted specifications; they do not establish
global optimization or filter stability.

\end{document}